%% file: main.tex
\documentclass[11pt]{article}

\usepackage[final]{acl}

\usepackage{times}
\usepackage{latexsym}

\usepackage[T1]{fontenc}

\usepackage[utf8]{inputenc}

\usepackage{microtype}

\usepackage{amsthm}
\usepackage{amsmath}
\usepackage{inconsolata}
\usepackage{booktabs}
\usepackage{cleveref}
\usepackage{multirow}
\usepackage{subcaption}
\usepackage{enumitem}
\usepackage{float}
\usepackage{soul}
\usepackage{graphicx}
\usepackage[colorinlistoftodos]{todonotes}
\usepackage{comment}

\newcommand{\inlineemoji}[1]{\includegraphics[height=1em]{emojis/#1.png}}
\usepackage{fontawesome5}
\sethlcolor{gray!20}

\newcommand{\CLS}{\texttt{[CLS]}}
\newcommand{\SEP}[1][]{\texttt{[SEP#1]}}
\newcommand{\hf}{\inlineemoji{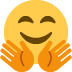}}

\newcommand{\githubrepo}[1]{%
    \begin{center}
        \href{#1}{\small \faGithub \ \texttt{#1}}
    \end{center}
}

\newtheoremstyle{compactpar}
  {3pt}   
  {3pt}   
  {\rmfamily}  
  {}      
  {\bfseries} 
  {}      
  {0.5em} 
  {#1 \thmnumber{#2}} 

\theoremstyle{compactpar}

\newtheorem{masking_step}{Masking Step} 
\crefformat{masking_step}{Mask #2#1#3}

\newtheorem{researchQ}{RQ} 
\crefformat{researchQ}{RQ #2#1#3}

\newcommand{\allowed}{\leftarrow}
\newcommand{\blocked}{\not\leftarrow}

\begin{document}


\title{MICE: Minimal Interaction Cross-Encoders for efficient Re-ranking}

\author{
  \textbf{Mathias Vast}$^{* 1,2}$ \quad
  \textbf{Victor Morand}$^{* 1}$ \quad
  \textbf{Josiane Mothe}$^{3}$ \quad
  \textbf{Benjamin Piwowarski}$^{1}$ \\[4pt]
  $^{1}$Sorbonne Universit\'e, CNRS, ISIR, Paris, France \\
  $^{2}$Sinequa by ChapsVision, Paris, France \\
  $^{3}$Univ. Toulouse, IRIT, CLLE, CNRS, Toulouse, France \\[6pt]
  \small \textbf{Correspondence:} \href{mailto:mathias.vast@isir.upmc.fr}{mathias.vast@isir.upmc.fr} \\ 
  \small $^{*}$Equal contribution.
}
\maketitle

\begin{abstract}

In Information Retrieval (IR), cross-encoders deliver state-of-the-art ranking effectiveness but have a high inference cost, limiting their use to second-stage re-rankers. 
Prior work has addressed this bottleneck from two largely separate directions: accelerating cross-encoder inference  through attention sparsification, or improving first-stage retrieval effectiveness to alleviate the need of a re-ranker, using more complex models, e.g. late-interactions.
In this work, we bridge these two directions through an in-depth analysis of cross-encoder internal mechanisms. By identifying and removing superfluous interactions, we derive MICE (Minimal Interaction Cross-Encoders), a new cross-encoder architecture that retains effectiveness while reducing computational overhead. Extensive evaluations show MICE retains most of the performances of its cross-encoder counterparts in-domain and matches or even exceeds it in out-of-domain, while reducing FLOPs down to $2.5$ times.
\end{abstract}

\githubrepo{https://github.com/xpmir/mice}

\section{Introduction}

Since the advent of transformers \cite{transformers}, many neural Information Retrieval (IR) architectures have been proposed, from representation-based bi-encoders to interaction-based cross-encoders, with dense or sparse, single- or multi-vector representations.
Although cross-encoders offer state-of-the-art ranking performance, they are computationally prohibitive when applied exhaustively to large corpora~\cite{yates2021pretrained,monobert}. This bottleneck has maintained the prevalent \textit{retrieve-and-rerank} paradigm, where a fast initial retriever \cite{karpukhin-etal-2020-dense,bm25} filters the corpus down to a candidate pool that is then re-ranked by a cross-encoder \cite{monobert}.

However, the \textit{retrieve-and-rerank} paradigm introduces fundamental limitations. First, re-ranking can only be as good as its input: documents missed by the first-stage retriever cannot be recovered. Second, the re-ranking step is still affected by the low efficiency of cross-encoders.

\begin{figure}[t]
    \centering
    \includegraphics[width=\linewidth]{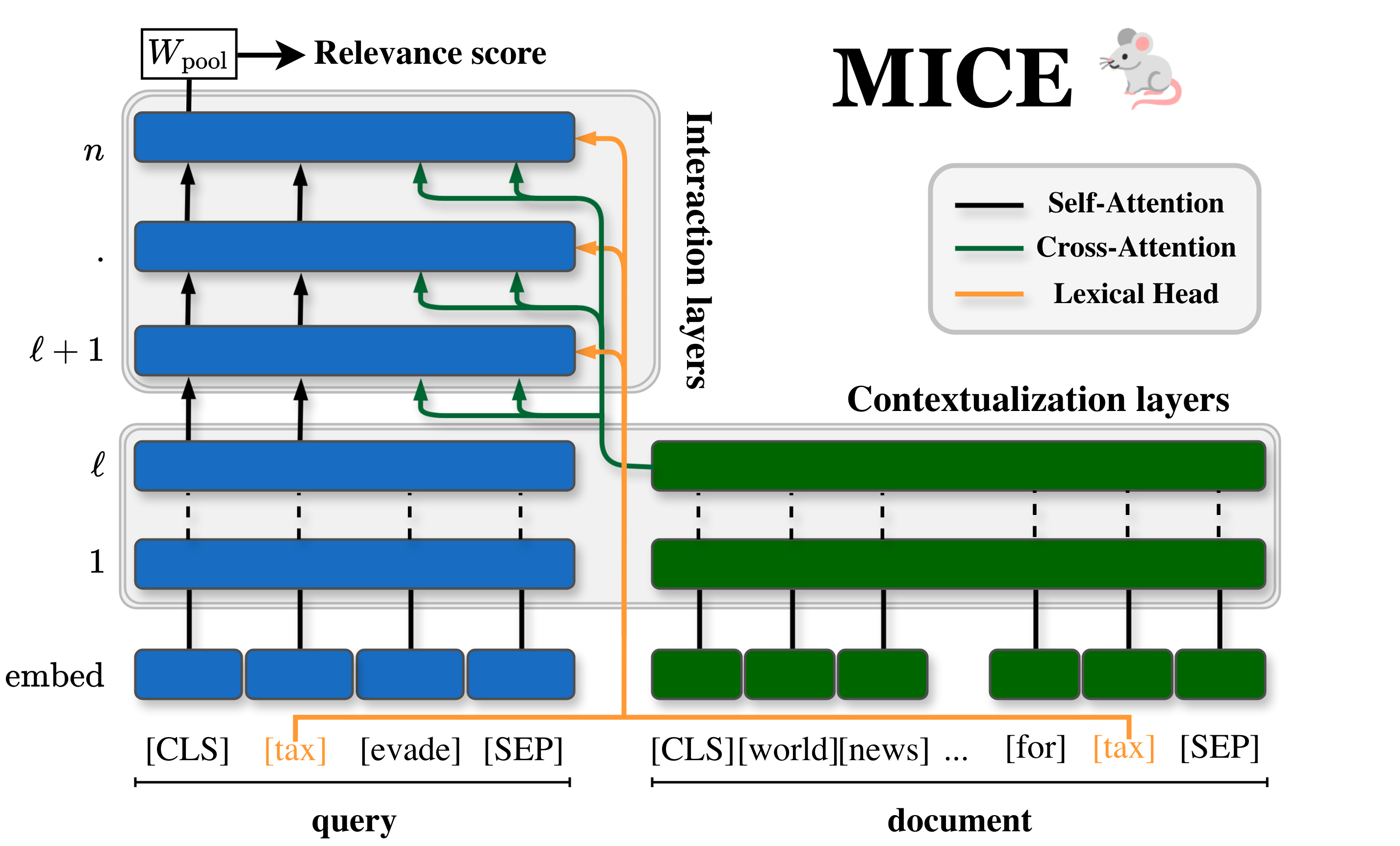}
    \vspace{-0.7cm}
    \caption{MICE architecture. Keeps the strict minimum interactions in a cross-encoder to maintain effectiveness.}
    \label{fig:MICE}
    \vspace{-0.5cm}
\end{figure}

The first limitation is addressed by architectures that improve effectiveness while remaining efficient enough for full-corpus search. 
A representative class of models are late-interaction models, such as ColBERT \cite{khattab2020colbert},  that keep a token level representation of documents and queries (like cross-encoders), but delay (only contextualized representations) and simplify (MaxSim) their interactions. Late-interaction models remain less effective than cross-encoders.
The second limitation has been approached by cutting cross-encoder inference cost by architectural optimizations such as sparse attention \cite{schlattInvestigatingEffectsSparse2024}, or by restricting the candidate pool via pruning / cascade strategies \cite{ranked_list_truncation,campagnano2025e2rank}. Still, such methods remain computationally heavier than late-interaction models.

Our work connects these two lines of research by deriving an efficient, late-interaction-style re-ranker directly from a conventional cross-encoder. Specifically, we pursue two main goals. 
First, we propose a principled masking strategy that strips unnecessary cross-encoder interactions identified by prior interpretability studies \cite{lu-etal-2025-pathway,zhan2020bertAnalysisReranker}, reducing inference cost to below that of ColBERT while preserving ranking effectiveness (\Cref{sec:masking}).
Second, we use these findings to reshape the cross-encoder architecture into a more efficient but as effective re-ranking model.
We call this new architecture MICE (Minimal-Interaction Cross-Encoder, described in \Cref{sec:MICE} and depicted in \Cref{fig:MICE}). We demonstrate that MICE does not degrade the ranking performance compared to the initial cross-encoder, while being up to $2.5\times$ more efficient, on two distinct backbones, BERT \cite{bert} and the more recent ModernBERT \cite{modernbert}.

We address the following research questions:

\begin{researchQ}\label{rq:TrainPruneAttn}
    How many interactions can be removed to improve the efficiency of a cross-encoder, while maintaining its effectiveness? 
\end{researchQ}


\begin{researchQ}\label{rq:NewArchi}
    Can we design a more efficient architecture, while maintaining cross-encoder effectiveness?
\end{researchQ}

\section{Related Works}

\textbf{More efficient ranking paradigms}
\noindent Cross-encoder models \cite{monobert} are  highly effective, but their poor efficiency has led IR practitioners to explore more efficient alternatives. Bi-encoders encode queries and documents separately into single vectors \cite{karpukhin-etal-2020-dense}.
This enables offline document indexing, making bi-encoders very efficient first-stage retrievers, at the cost of reduced effectiveness, particularly in out-of-domain (OOD) scenarios \cite{rosa2022defensecrossencoderszeroshotretrieval,thakur_beir_2021}, as the model cannot explicitly capture query-document interactions. 
Learned Sparse Retrieval models, such as SPLADE \cite{formal2021splade}, improve the effectiveness of dense bi-encoders while retaining their efficiency due to their sparse nature and inverted indexes.

In contrast with bi-encoders, late-interaction models, such as ColBERT \cite{khattab2020colbert},  encode queries and passages into multiple vectors (one per token) and aggregate the score of each query token by computing the maximum similarity (\textit{MaxSim} operator) with a document token. This yields greater effectiveness at the cost of increased storage, partially addressed by follow-up works \cite{santhanam2022colbertv2,santhanam2022plaid}. 

Several hybrid architectures have been proposed to bridge the efficiency-effectiveness gap between bi-encoders and cross-encoders. Poly-encoders \cite{Humeau2020Poly-encoders} summarize documents into $M$ context vectors to reduce self-attention cost, but only match cross-encoder effectiveness at high $M$ values ($M=360$, exceeding the average MS MARCO document length). 

 Mid-fusion transformers \cite{tan-bansal-2019-lxmert} encode each input stream independently in the lower layers before fusing their representations in the upper layers, an idea later adapted to text pairs by DeFormer \cite{caoDeFormerDecomposingPretrained2020} to enable offline passage encoding.
 PreTTR \cite{macavaneyEfficientDocumentReRanking2020} apply this idea to IR, while MORES \cite{gaoModularizedTransfomerbasedRanking2020mores} further disentangle document and query processing into two distinct encoders before combining them with an interaction module.
This design enables offline document encoding, but hybrid architectures struggle to match the effectiveness of cross-encoders.

Instead of modifying the cross-encoder paradigm, several works have aimed at improving the efficiency of re-rankers, without altering their fundamental architecture.

\noindent \textbf{Improving the efficiency of cross-encoders}
 A first line of research targets the self-attention mechanism \cite{attention}, which underpins transformer representations but slows inference. General approaches such as linear attention \cite{wang2020linformer,Wu2021FastformerAA} have been proposed to reduce its complexity, yet they generally transfer poorly to IR, where attention plays a central role in detecting semantic and lexical matches between query and document tokens \cite{lu-etal-2025-pathway}. Sparse attention models such as Longformer \cite{Beltagy2020LongformerTL} and BigBird \cite{Bigbird} restrict token interactions to local windows, under the assumption that not all pairwise interactions are necessary. \cite{schlattInvestigatingEffectsSparse2024} successfully applied this principle to cross-encoders with limited effectiveness loss. A second line of research focuses on reducing model size through knowledge distillation \cite{hinton2015distillingknowledgeneuralnetwork} or pruning \cite{lottery_ticket,campos2023sparsebertsparsemodelsgeneralize}, 
 with applications to IR rankers \cite{lei2025makinglargelanguagemodels,Schlatt_2025}.

Rather than optimizing existing cross-encoders or replacing them entirely, our methodology progressively strips cross-encoders of unnecessary interactions to improve their efficiency (see \autoref{sec:masking}). Pushed to its limit, this process naturally leads to a new architecture, namely MICE (see \Cref{sec:MICE}).

\section{Towards minimal interaction cross-encoders }\label{sec:masking}

In this section, we first study which interactions are truly necessary within a cross-encoder by analyzing the impact of masking interactions between input segments: \CLS{}, query $Q$, document $D$, and \SEP{} tokens.

\subsection{Background}

Cross-encoders classify a query-document $(Q,D)$ couple. Their input is typically composed of the sequence of tokens: \texttt{[CLS] $q_1 \ldots q_n$ [SEP1] $d_1 \ldots d_m$ [SEP2]} ($n$ query tokens and $m$ document tokens).
In this work we focus on encoder-only cross-encoders, as this architecture has been shown to be the best choice for sequence classification task \cite{ettin}.
The relevance score for the document \textit{w.r.t.} the query is predicted by applying a classification head to the token representation \CLS{} of the last transformer layer $L$. 

Cross-encoders use self-attention to capture \emph{interaction} signals between two tokens $a$ and $b$. 
In IR, these interactions have specific meaning when token $a$ belongs to the query and token $b$ to the document (or reciprocally). By moving information from token $a$ and comparing it with what is already encoded in token $b$ (its identity, the semantics of its context, etc.), self-attention is key to detecting \emph{matching signals} between queries and documents.

For that reason, studies attempting to reverse-engineer the inner working of cross-encoders particularly focus on  interpreting the interactions between the different input parts through the self-attention. 
For instance, \citet{lu-etal-2025-pathway} show that it allows cross-encoders to detect, not only exact matching signals -- what a lexical retriever like BM25 \cite{bm25} would do -- but also semantic matching signals.
\citet{zhan2020bertAnalysisReranker} show that the relevance prediction process inside cross-encoders is decomposed into multiple consecutive stages. In the first layers, the model contextualizes query and document tokens. At this stage, query-document interactions only play a minor role, but once their semantics have been properly encoded, the model starts using query-document interaction signals. \citet{lu-etal-2025-pathway} further provide empirical evidence that matching signals are captured by the self-attention, in so-called \textit{matching heads}, and then aggregated inside the query tokens by \textit{contextual query representation heads}. Ultimately, \textit{relevance scoring heads} scan query tokens to retrieve relevance information and to encode it in the \CLS{} representation for the final prediction. 
Their findings suggest that information does not flow freely between input parts inside cross-encoders, but instead roughly flows from the document tokens towards the query tokens (after contextualization), and then towards the \CLS{} \cite{zhan2020bertAnalysisReranker}. 
In the following, we denote input parts as $X,Y \in \{\CLS, Q,$ \SEP[1], $D$, \SEP[2]\} and $Y \allowed X$ (resp.  $Y \blocked X$) a transfer of information (resp. blocking the transfer) from a token in $X$ \emph{to} a token in $Y$ through the self-attention mechanism (or equivalently that $Y$ \emph{attends to} $X$).

\subsection{Our approach}

\begin{figure}[t]
    \centering
  \includegraphics[width=0.9\linewidth]{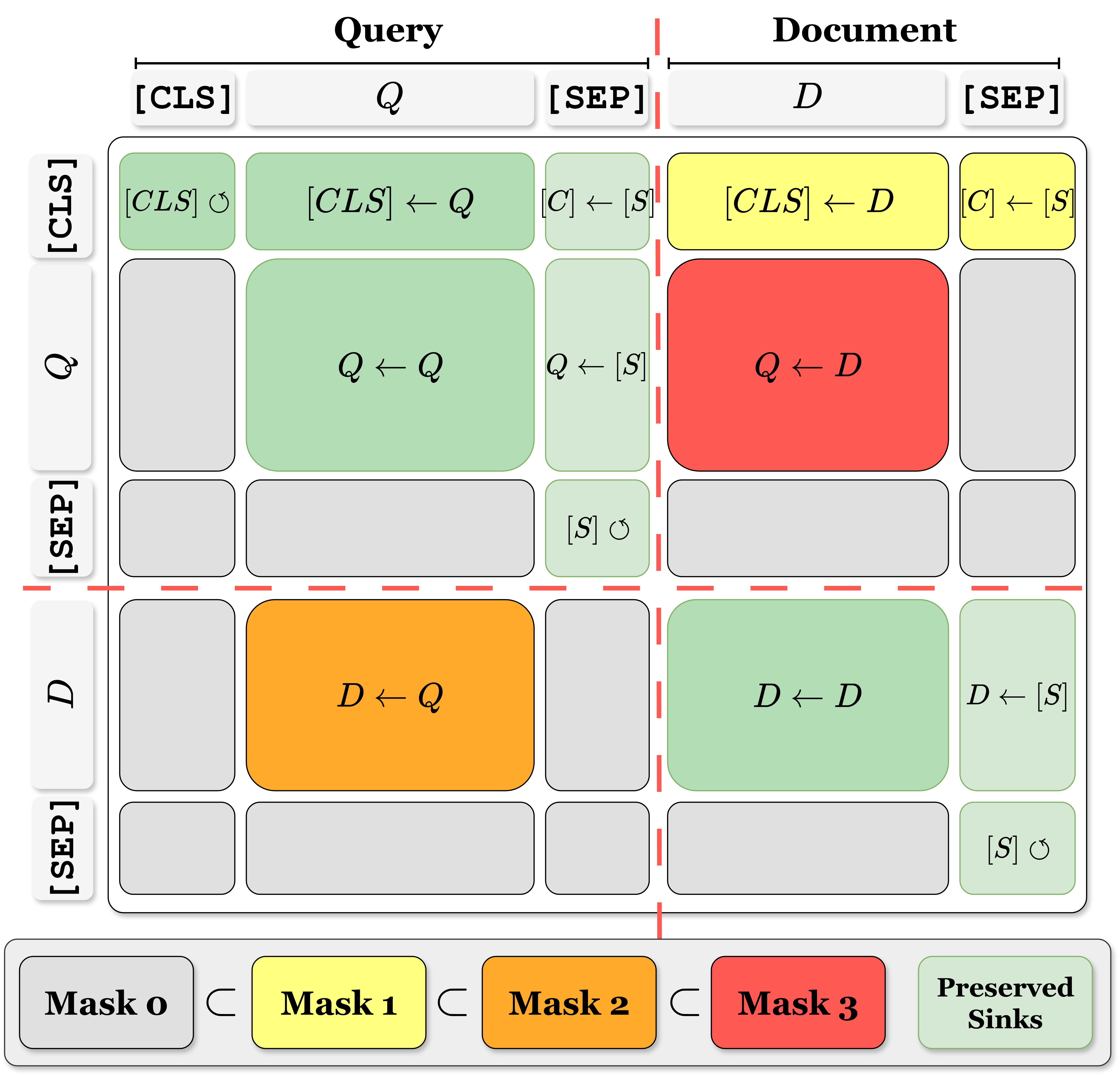}
    \vspace{-0.3cm}
    \caption{Masking approach.  Interactions between input parts (\CLS{}, $Q$, $D$, \SEP{}) are blocked using cumulative masking. Colors indicate the step where masking begins, ending with \Cref{mask_step:3} in a complete $Q\not \leftrightarrow D$ separation (block-diagonal structure). Green blocks denote permanently preserved interactions and attention sinks.}\label{fig:first_step}
    \vspace{-0.5cm}
\end{figure}

We measure the importance of a given interaction with causal analysis: by evaluating the effect of blocking it on the model's effectiveness. In practice, we prevent interactions by \textit{masking} the corresponding block in the self-attention weight matrices, i.e., setting its logits to $-\infty$ before \textit{Softmax}, effectively blocking any information transfer. 
Following prior description of information flow between input parts in cross-encoders \cite{zhan2020bertAnalysisReranker,lu-etal-2025-pathway}, we consider four cumulative masking steps, all summarized in \Cref{fig:first_step}. For instance, \Cref{mask_step:2} blocks information transfers from the query $Q$ to the document $D$ ($D \blocked Q$), while also applying \Cref{mask_step:0} and \Cref{mask_step:1}.

\begin{masking_step} \label{mask_step:0}
We block all interactions towards \SEP{} and from \CLS{} to other input parts, while dedicating \SEP[1] and \SEP[2] as attention sinks respectively for $Q$ and $D$. This step is expected to have little to no impact on effectiveness, as it merely reduces noise received by query and document tokens.
\textit{\cref{mask_step:0}: $\SEP \blocked \{\CLS,Q,D\}$, $\{Q,\SEP,D\} \blocked \CLS$, $Q\blocked$\SEP[2] and $D\blocked$\SEP[1]}.
\end{masking_step}

\begin{masking_step} \label{mask_step:1}
We block the flow of information from the document to \CLS{} ($\CLS \blocked \{D$,\SEP[2]$\}$), motivated by evidence that relevance signals are stored in query tokens before being passed to \CLS{} \cite{lu-etal-2025-pathway}. This step is expected to have only marginal impact on effectiveness.
\textit{ \Cref{mask_step:1}: \cref{mask_step:0} and $\CLS \blocked \{D$,\SEP[2]$\}$}.
\end{masking_step}

\begin{masking_step} \label{mask_step:2}
We mask the query-to-document flow ($D\blocked Q$) across all layers, with limited expected effectiveness drop. This is a first step towards separating query and document contextualization, enabling offline document encoding.
\textit{\cref{mask_step:2}: \cref{mask_step:1} and $D\blocked Q$}
\end{masking_step}

\begin{masking_step} \label{mask_step:3}
We additionally block document-to-query interactions ($Q\blocked D$) across the first $\ell^*$ layers, fully separating query and document contextualization in the early layers. $\ell^*$ is the highest value that preserves the base model's effectiveness.
\textit{\Cref{mask_step:3}: \cref{mask_step:2} and $Q\blocked D$ (up to layer $\ell^*$)}
\end{masking_step}

\noindent
Appendix~\ref{appendix:mask_details} provides further rationales behind these masking steps.
With this approach, the set of possible interactions decreases after each step, until we obtain the minimal set required to maintain the original cross-encoder effectiveness. 

\subsection{Experimental setup}\label{sec:method}

\paragraph{Backbones}

We consider two distinct backbones: BERT \cite{bert}, which has been extensively studied \cite{rogers-etal-2020-primer,ferrando2024primerinnerworkingstransformerbased}
, and ModernBERT \cite{modernbert}, a recent update to the original BERT architecture with stronger capabilities.
While our approach can be applied to analyze any cross-encoder, we focus here on cross-encoders based on small transformer encoders: MiniLM-v2 \cite{minilmv2} (simply "MiniLM" in the paper), a compact yet effective model based on BERT \cite{bert} optimized via deep self-attention distillation, and Ettin-32M \cite{ettin}, based on ModernBERT \cite{modernbert}.
We further detail their configuration in Appendix (\Cref{tab:model-configs}). 
Small models let us run the extensive training and evaluation of our masking study, which would be prohibitively expensive on larger backbones. Despite the rise of LLMs in IR \cite{rankLlama}, they remain competitive for ranking thanks to their efficiency \cite{Dejean2024ATC}.

\paragraph{Baselines} 


We consider two baselines. Sparse CE \citep{schlattInvestigatingEffectsSparse2024} that sparsifies cross-encoder attention by blocking document-to-query interactions ($Q \blocked D$), the direction identified as most important by previous work \cite{zhan2020bertAnalysisReranker,lu-etal-2025-pathway}, yet preserves effectiveness when fine-tuned with this mask, making it a strong reference for validating our design choices. 
We do not reproduce its sliding-window attention over document tokens, as this is orthogonal to our work and detrimental for small windows. 
We also compare with PreTTR \cite{macavaneyEfficientDocumentReRanking2020}, which implements mid-fusion by blocking all query-document interactions in the early layers, differing from our approach in the later layers where it keeps all interactions. 
We only reproduce its independent query-document contextualization across the first $\ell$ layers, without the Auto-Encoder compression — reported results thus constitute an upper bound for PreTTR.

\subsection{Results and Analysis}

\input{new_figures/Table1}

We fine-tune pretrained models with the masks on the re-ranking task (detailed setup in Appendix \ref{appendix:experimental_framework}), learning a separate cross-encoder for each mask and for the baselines of Section \ref{sec:method}. \Cref{tab:MskEvalsFT} reports the average nDCG@10 on 5 random seeds across ID — MS MARCO \cite{msmarco}, TREC-DL19 and 20 \cite{craswell2019overview,craswell2020overview}—  and OOD —BEIR \cite{thakur_beir_2021}— described in Appendix \ref{appendix:datasets}.


First, we note that fine-tuning these checkpoints with \Cref{mask_step:0}, either matches the unmasked baselines' effectiveness (for Ettin) or exceeds them (for MiniLM, especially in OOD: 46.3 vs 44.7 nDCG@10), 
confirming that the role of the \SEP{} tokens is not directly tied to relevance prediction. 

For both backbones, \Cref{tab:MskEvalsFT} shows that further masking up to \Cref{mask_step:2} either exceeds or matches the performance of the fine-tuned baselines, both ID and OOD ($+5.5$ nDCG@10 in average on OOD for \Cref{mask_step:1} on MiniLM). With this masking strategy, MiniLM reaches the best performances, both for ID and OOD. 
We can draw two conclusions from that:
(i) it confirms that \CLS{} does not need to attend to the document to receive the appropriate relevance signals (\cref{mask_step:1}). In contrast, OOD results suggest that cross-encoders capture \emph{spurious} correlations when allowing this transfer of information;
%
and (ii) Using \cref{mask_step:2} ($D \blocked Q$) across all layers does not harm cross-encoders effectiveness, while it potentially enables substantial gain in efficiency. 
This also contradicts the masking strategy of Sparse CE \cite{schlattInvestigatingEffectsSparse2024} ($Q \blocked D$) as we observe no improvement on the effectiveness compared to the unmasked MiniLM cross-encoder when applying their sparsification. 
This strengthens our design choice for simultaneously improving both effectiveness and efficiency.

\begin{figure}[t]
    \centering
    \includegraphics[width=\linewidth]{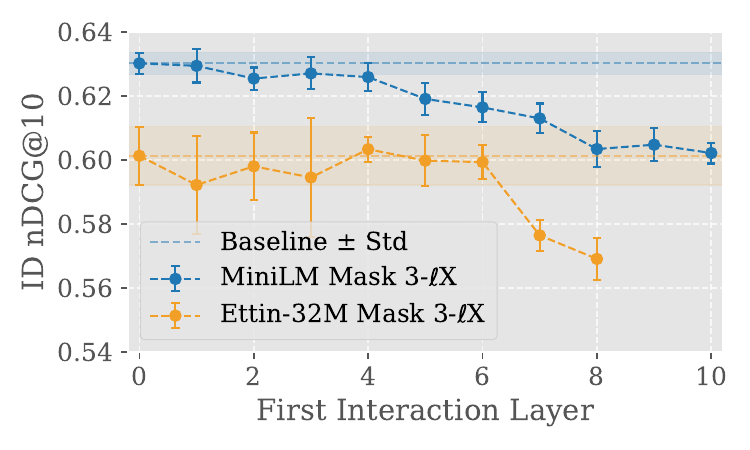}
    \caption{Search of $\ell^*$ (\cref{mask_step:3}). \textit{All interactions between $Q$ and $D$} up to a given layer are masked. } \label{fig:layer_merge}
    \vspace{-0.4cm}
\end{figure}

\Cref{fig:layer_merge} shows the  nDCG@10 averaged on ID collections as a function of the layer $\ell$ before which query and document tokens can be processed independently, \textit{i.e.}, without interaction (\cref{mask_step:3}).
For MiniLM, we observe that it is possible to contextualize the query and document tokens independently, without harming ID effectiveness, up to the layer $\ell^*=4$ (out of 12 total layers). There are two potential explanations for the ID effectiveness drop after $\ell\geq 5$: (1) the remaining \textit{interaction layers} are not enough to properly capture relevance; (2) after layer $\ell^*=4$, further contextualizing query and document tokens independently introduces signal detrimental to the IR task. We partially address this question in \Cref{sec:MICE}, when designing MICE's architecture. For Ettin-32M, the optimal $\ell^*$ is 6 (out of 10). We refer to $\ell^*$ as the \emph{first interaction layer} for each model.

\autoref{tab:MskEvalsFT} further reports the detailed results of \cref{mask_step:3} with both models, using their respective layer $\ell^*$. We observe a consistent drop in OOD performance compared to using only \cref{mask_step:2}, in particular for Ettin-32M, with almost -4 for nDCG@10. 
At the same time, we observe very little variation in ID effectiveness between masking steps, and compared to the baseline (see the evolution, per backbone, in the ID column of \autoref{tab:MskEvalsFT}). This suggests that our masks do not prevent the models from learning correctly the IR task, but, depending on the base model, may hinder their OOD robustness. 
%

We also report a PreTTR reproduction on MiniLM using $\ell^* = 4$, which performs similarly to \Cref{mask_step:2} and surpasses \Cref{mask_step:3} by 1--2 points on average over ID and OOD. This gap stems from the richer interactions allowed in PreTTR's later layers. Still we expect our additional masks to yield a better efficiency-effectiveness trade-off overall.
 

\noindent\textbf{Intermediate Conclusions}\label{sec:MaskingCCL}
Our experiments with \Cref{mask_step:1} first revealed that masking direct $\CLS \blocked D$ interactions consistently improves both ID and OOD effectiveness. 
Then, \Cref{mask_step:2} confirmed that information can flow only from the document to the query ($Q \allowed D$), and  blocking $D \blocked Q$ does not harm the model. 
Finally, \Cref{mask_step:3} corroborated previous studies on mid-fusion architectures \cite{macavaneyEfficientDocumentReRanking2020, caoDeFormerDecomposingPretrained2020}, showing that queries and documents can be contextualized independently in the lower layers, without compromising the performance.
This section demonstrates that masking targeted interactions in a cross-encoder can enhance its re-ranking performance, addressing \Cref{rq:TrainPruneAttn}. This gain is particularly pronounced in OOD, suggesting that our masks act as a form of regularization, preventing the model from overfitting. 
This observation holds across backbones and sets the stage to more efficient and as effective cross-encoders.


\section{MICE - Minimal Interaction Cross Encoder}\label{sec:MICE}

Building on the insights from \Cref{sec:masking}, we propose a novel, streamlined late-interaction style ranker architecture we call MICE (\textbf{M}inimal \textbf{I}nteraction \textbf{C}ross-\textbf{E}ncoder).
By discarding superfluous interactions, we show that MICE becomes up to 2.5$\times$ more efficient than standard cross-encoders, while maintaining competitive re-ranking performance (\Cref{rq:NewArchi}).

\subsection{Architecture}\label{sec:archi}
Compared to a standard cross-encoder, MICE differs in four principal architectural choices: 
\textbf{(1) Mid-Fusion} first encodes the query and document independently;
\textbf{(2) Light Cross-Attention} only transfers information from a frozen document representation to the query;
\textbf{(3) Layer Dropping} reduces the number of interaction layers; 
\textbf{(4) Lexical Head} optionally transfers lexical (exact match) information to upper interaction layers.
We detail each of these choices in the following paragraphs, and provide an overview of MICE in \Cref{fig:MICE}.

\begin{description}[leftmargin=0pt, style=sameline, noitemsep]
\item[Mid-Fusion]
First, MICE follows the Mid-Fusion paradigm by encoding query and document tokens independently. We use the optimal fusion layer $\ell^*$ derived from our \Cref{mask_step:3} analysis (\Cref{sec:MaskingCCL}) to set the number of contextualization layers. 
MORES~\cite{gaoModularizedTransfomerbasedRanking2020mores} pushes this further by dedicating two entirely separate transformer models to document and query encoding, at the cost of efficiency.

\item[Light Cross-Attention]
As \Cref{mask_step:2} suggests that $D \blocked Q$ direction is superfluous, we go further by only computing cross-attention from \emph{frozen document representations} to the query, eliminating expensive self-attention over document tokens. This significantly differs from PreTTR, which retains full query-document attention in its interaction layers, and is close to the interaction module of MORES~\cite{gaoModularizedTransfomerbasedRanking2020mores} — though our masking analysis provides the first empirical justification. 
Despite sharing similarities with late-interaction architectures like ColBERT \cite{khattab2020colbert}, the cross-attention offers higher expressivity as MICE can capture more granular, non-linear dependencies between query and document tokens. Following MORES, we compute cross-attention before self-attention in the transformer layer, allowing query tokens to gather document information first. We also allow cross-attention transfers from document to \CLS{} (as opposed to \Cref{mask_step:1}), a choice validated by ablations in \Cref{app:Ablations}. Cross-attention is seeded with the original self-attention weights. 

\item[Dropping Backbone Top Layers]
Unique to MICE, we further posit that the final layers of a backbone pretrained on MLM may be non-essential for re-ranking, as specialized for token prediction. Consequently, we experiment dropping them to limit the number of \textit{interaction layers}.

\item[Lexical Head]
Experiments with \Cref{mask_step:3} showed that separately contextualizing $Q$ and $D$ beyond $\ell^*$ is detrimental (Fig. \ref{fig:layer_merge}). Overly semantic representations may hinder \emph{lexical matching signals}. We thus also introduce a novel dedicated attention head in the interaction layers that explicitly leverages exact query-document token matches. We implement it as an additional cross-attention head that replaces the standard query-key dot product with precomputed lexical match scores (for query/document tokens $(q_i, d_j)$, $s_{i,j} = 1 \text{ iif } q_i = d_j$ or $s_{i,i}=1$ if $\not \exists d_j | \ q_i = d_j$ to ensure normalization ). It is thus parametrized only by the value/output matrices.
\end{description}

\paragraph{Theoretical Efficiency Gains}
These architectural choices offer significant efficiency gains over a full cross-encoder.
(i) Queries and documents are processed independently in contextualization layers, reducing the cross-encoder quadratic cost of query-document attention on concatenated sequence. (ii) In upper interaction layers, document representations remain frozen, skipping expensive MLP and self-attention computations for document tokens and updating only query tokens with cross-attention to the document. This yields massive savings since documents are much longer than queries. Finally, (iii) layer pruning discards unnecessary transformer blocks entirely, directly reducing total network depth and overall compute load.
We provide quantitative comparisons by computing an estimate of the FLOPs necessary to produce one score for each configuration of MICE. 

\subsection{Experimental Setup}

\noindent \textbf{Baselines} In addition to comparing MICE variants to their corresponding unmasked cross-encoders (same baselines as in \Cref{sec:masking}), we also compare MICE to previous hybrid architectures, i.e., PreTTR \cite{macavaneyEfficientDocumentReRanking2020} and MORES \cite{gaoModularizedTransfomerbasedRanking2020mores}. 
As MICE resembles late-interaction like models, we further compare it with a reproduction of \textbf{ColBERT} \cite{khattab2020colbert}.
We train such approaches from MiniLM-v2, fine-tuning them with the same setup as our MICE models (described in Appendix \ref{appendix:experimental_framework}, ensuring a fair comparison).

\noindent\textbf{Validation Set} Our training setup for MICE slightly differs from this of the masking experiments, as we validate on nano-BEIR \cite{NanoBEIRMultilingualInformation2026} instead of MS MARCO. 
We view nano-BEIR as a richer tool to select checkpoints, while it allows us to still use BEIR as an "OOD" benchmark, given the very limited size of its subsets. On the contrary, to avoid biasing the architecture for BEIR, we chose to validate all the design choices of MICE on ID collections only. This also includes the masking experiments and the validation on MS MARCO.



\subsection{Results}\label{sec:MICE_results}

\begin{figure}[t]
    \centering
    \begin{subfigure}[t]{0.45\textwidth}
        \centering
        \includegraphics[width=\textwidth]{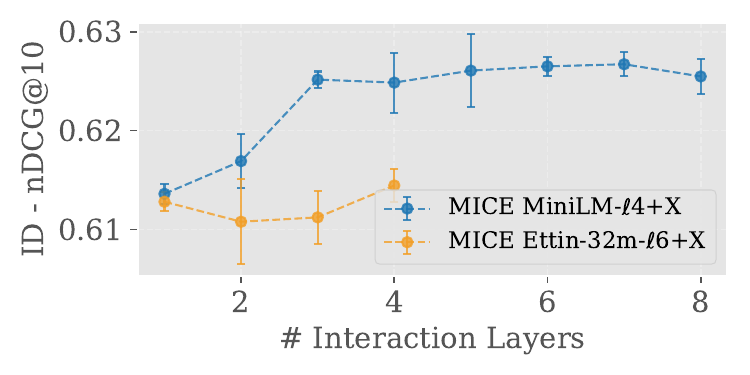}
    \end{subfigure}
    \vfill
    \vspace{-0.2cm}
    \begin{subfigure}[t]{0.45\textwidth}
        \centering
        \includegraphics[width=\textwidth]{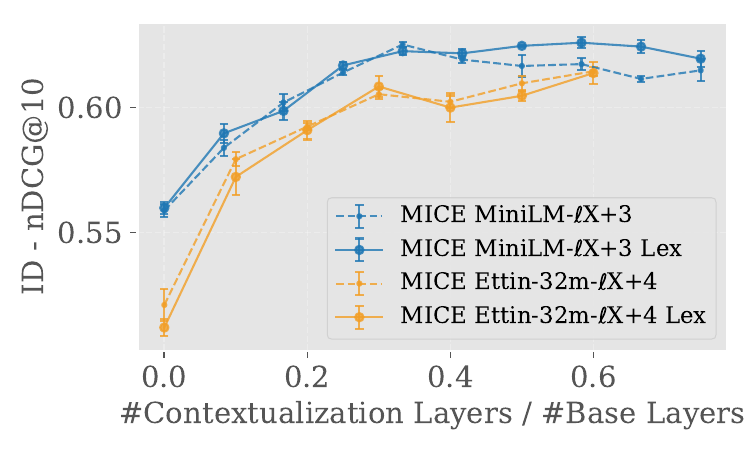}
    \end{subfigure}

    \caption{Impact of number of interaction / contextualization layers and lexical head (``Lex'' models) in MICE (mean ID effectiveness over 3 seeds).} \label{fig:MiceGridSearch}
    \vspace{-0.3cm}
\end{figure}

For MiniLM and Ettin-32M backbones, we first analyze in \Cref{fig:MiceGridSearch} the impact of the number of contextualization and interaction layers kept from the backbone. 
We report detailed evaluations of MICE variants in \Cref{tab:MICE-evals}, including a FLOPs-based theoretical speedup over the full cross-encoder (the `$\times$' column, computed per forward pass without pre-computing documents; see Appendix \ref{app:flops}): in practice, we compute it as FLOPs(baseline)  / FLOPs(model) to get the approximate speedup for computing one relevance score. 


\input{new_figures/TableMicev2}

\noindent \textbf{Layer Dropping} As shown in \Cref{fig:MiceGridSearch} (top) and \Cref{tab:MICE-evals}, dropping late backbone layers does not hurt — and can even benefit — re-ranking effectiveness, both ID and OOD. MiniLM plateaus at 3 interaction layers (MICE-$\ell$4+3 outperforms MICE-$\ell$4+8 despite using only 7 of 12 layers), while Ettin-32M shows higher variance, making conclusions harder to draw. MICE-all variants show no statistically significant difference from those using fewer interaction layers.

\noindent \textbf{Contextualization layers} \Cref{fig:MiceGridSearch} (bottom) shows ID effectiveness with increasing number of contextualization layers. For both backbones, effectiveness plateaus at $\sim$half of the backbone layers, even decreasing after $\ell^*=4$ for MiniLM without lexical heads.
It corroborates the optimal contextualization depths identified in \Cref{fig:layer_merge} ($\ell^*=4$ for MiniLM, 6 for Ettin-32M). \Cref{tab:MICE-evals} provides further evidence, as MICE-MiniLM-$\ell$4+3 reaches an nDCG@10 of 49.7 on BEIR (+5.7 points over its cross-encoder counterpart), while underperforming in ID. At the same time, MICE-MiniLM-$\ell$4+3 achieves better ID and OOD results than both PreTTR and MORES. Ettin-32M derived MICE models show a more modest trend, slightly underperforming their baseline (47.2 for MICE-$\ell$6+4 vs. 48.4).

\noindent \textbf{Lexical Head} Deeper layers provide richer semantic signals but at the cost of lexical information, whose absence may explain the effectiveness drop beyond optimal $\ell^*$ values (\Cref{fig:layer_merge}). Our lexical head is designed to alleviate this trade-off, allowing deeper contextualization without sacrificing lexical matching.
%
The ``Lex'' variant recovers the effectiveness drop beyond $\ell^*=4$ for MiniLM (\Cref{fig:MiceGridSearch}), confirming that the lexical head preserves lexical information in deeper variants. No comparable gains are observed for Ettin-32M, highlighting architectural differences between BERT~\cite{bert} and ModernBERT~\cite{modernbert}. \Cref{tab:MICE-evals} confirms this: MICE-MiniLM-$\ell$8+3 Lex outperforms its default counterpart in ID and achieves the best OOD results overall (50.5).



\noindent \textbf{Baselines Comparisons} Regarding effectiveness, \Cref{tab:MICE-evals} shows that our MICE-MiniLM variants can outperform our MiniLM-based reproduction of PreTTR and MORES in OOD. In ID, while MICE-MiniLM-$\ell$8+3 slightly underperforms, the addition of the Lexical Head successfully fills the gap. However, MICE’s primary advantage lies in its efficiency: MICE-$\ell$4+3 is 2.5$\times$ more efficient than a full cross-encoder. In contrast, PreTTR and MORES do not reduce computational overhead (FLOPs), as inputs must still traverse all transformer layers.
Comparisons with our reproduced MiniLM-ColBERT on ID and BEIR validate that MICE enables richer interactions.

Appendices \ref{app:Ablations} and \ref{appendix:full_mice_results} detail some ablations justifying the final MICE architecture, as well as a more exhaustive \Cref{tab:MICE-evals} with detailed results of the BEIR datasets \cite{thakur_beir_2021}.

In summary, we showed that, despite being up to 2.5$\times$ lighter than a standard cross-encoder, MICE preserves most of the ID performance of a standard cross-encoder while improving OOD generalization.
The answer to \Cref{rq:NewArchi} is therefore positive – we successfully leveraged our masking analysis to transpose the effectiveness of a cross-encoder into a novel and more efficient architecture that eliminates superfluous interactions.

\subsection{Pareto Frontier with MICE}

To show the applicability of MICE beyond MiniLM and Ettin-32M, we now extend our initial study to new backbones. We use $\text{ELECTRA}_\text{base}$, a 110M parameter encoder \cite{clark2020electra} to extend MiniLM, and the 68M and 150M versions from the Ettin-suite \cite{ettin}.
We build MICE models following the guidelines from the \Cref{sec:MICE_results}: setting $\ell^*$ to $\sim$half the backbone's total layers, with 3 interaction layers for BERT-based encoders and 4 for ModernBERT.
We report ColBERTv2 results \cite{santhanam2022colbertv2} instead of our reproduction for better comparison. 

\begin{figure}[t]
    \centering
    \includegraphics[width=\linewidth]{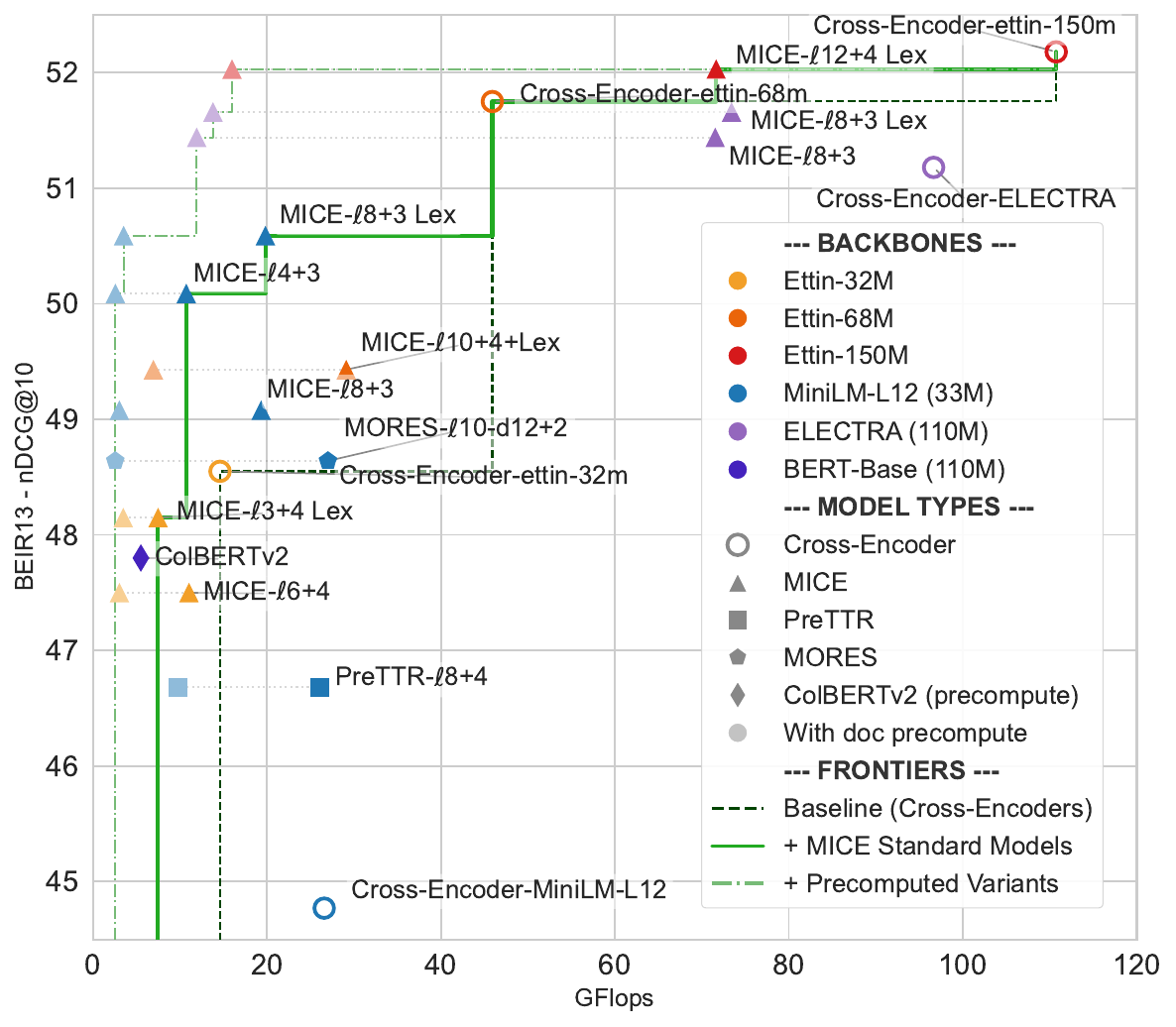}
    \vspace{-0.9cm}
    \caption{Pareto frontiers of all the models in our study, decomposed by backbones and types.} \label{fig:pareto}
    \vspace{-0.2cm}
\end{figure}

\noindent \textbf{Precomputing Representations} A key feature of MICE's Mid-Fusion architecture is the ability to pre-compute document representations offline. At inference, MICE only needs to encode the query, gather the document representations, and run both through its small set of interaction layers.
We present in \Cref{fig:pareto} an efficiency-effectiveness benchmark for all re-rankers considered in this work, featuring three Pareto frontiers: 1) baselines and standard cross-encoders; 2) the improvement brought by MICE; 3) a further frontier obtained by pre-computing document representations when possible (MICE, PreTTR, MORES, and ColBERTv2).

\Cref{fig:pareto} clearly shows efficiency gains from switching to MICE across all backbones. However, only BERT-based variants (MiniLM-v2 and ELECTRA) improve OOD effectiveness over their cross-encoder baselines and reproductions of PreTTR and MORES. On the Ettin suite, all variants except Ettin-150M show noticeable effectiveness drops, though efficiency gains still pushes the Pareto frontier compared to their cross-encoder counterparts. 

When pre-computing document representations, all Mid-Fusion models and ColBERT gain substantially in efficiency over traditional cross-encoders (e.g., MICE-Ettin-150M matches the efficiency of Ettin-32M while largely exceeding its OOD effectiveness). Overall, \Cref{fig:pareto} confirms MICE's generalization across architectures and scalability to larger models. While MICE-Ettin-68M slightly underperforms its standard counterpart, it remains more effective than Ettin-32M and can be more efficient via pre-computation, making it a valuable option for low-latency scenarios. We further discuss latency comparisons against ColBERTv2, with and without pre-computation, in the next section.

\section{Efficiency Analysis} \label{app:efficiency}

To complement the evaluation of the re-ranking effectiveness of MICE and of its efficiency in terms of FLOPs, we compare its latency and memory footprint with the standard cross-encoder architecture (entire forward pass over a query-document pair) and ColBERT (separately encoding the query and document before computing a \texttt{MaxSim} over their compressed representations). We present efficiency measures with and without precomputing the document representations, as ColBERT and MICE offer this possibility, contrary to a cross-encoder.
In practice, we use the same \texttt{MiniLM-L12-v2} backbone for each architecture (cross-encoder, ColBERT and MICE) to ensure a  fair comparison. We measure the inference time averaged over 100 forward passes on a maximum load setup (512 doc. token length, batch size 128 using an 12Gb Nvidia TITAN-V GPU) and report results in \Cref{tab:efficiency}. 

\input{new_figures/table_eff}

MICE achieves a $2\times$ speedup over standard cross-encoders, rising to $4\times$ with pre-computed document representations—effectively matching ColBERT's latency ($1.15\times$) while delivering superior performance (See \Cref{sec:MICE_results}). 
This is achieved by using fewer layers (i.e., the last layers of the backbone are dropped) and reducing interaction to cross-attention. Compared to ColBERT, we, however, have a larger memory footprint ($1.8\times$). This further confirms our conclusions drawn from the analysis of \Cref{fig:pareto}.

\section{Discussion}


\textbf{On the role of the lexical head} As noted in \Cref{sec:MICE}, we believe that the lexical head helps MICE models by restoring an "exact-match" signal that would otherwise be lost in the interaction layers. Our hypothesis: the greater the number of layers encoding independently the query and document tokens, the more likely that information useful for cross-encoders' BM25-like behavior—such as IDF statistics \cite{lu-etal-2025-pathway}—gets overwritten by "semantic" information from contextualization. The curves in \Cref{fig:MiceGridSearch} (bottom) for MiniLM-based MICE models support this statement: effectiveness drops passed a limit amount of contextualization layers (4), suggesting that an information is indeed lost. This result does not hold for Ettin-based MICE variants. Though our lexical head implementation is fairly simple, it has the merit of explicitly feeding exact-match signals into the interaction layers of the "MICE+Lex" variants.

\noindent \textbf{On the dependence to the backbone's architecture} MICE variants behave differently depending on the backbone: BERT-like models (MiniLM, ELECTRA, BERT-base) show slight ID effectiveness drops but clear OOD gains compared to their cross-encoder baselines. In contrast, ModernBERT shows drops in OOD as well. Since our architectural changes stem from interpretability studies conducted on BERT-like models \cite{zhan2020bertAnalysisReranker,lu-etal-2025-pathway}, this suggests ModernBERT-based cross-encoders behave differently and deserve their own interpretability analysis.

Importantly, MICE consistently improves the efficiency-effectiveness trade-off (\Cref{fig:pareto}), even when it does not improve effectiveness, making it a valuable addition to any IR pipeline.

\noindent \textbf{Implications for future works} MICE narrows the gap between cross-encoder effectiveness and late-interaction efficiency. Future work could further improve efficiency, building on methods that compress ColBERT-style token vector storage \cite{chirkova2025provence,szilvasy2026self}, by reducing token count and/or vector dimensionality. Another direction is refining MICE's interaction layers to bring more of the cross-encoder's modeling power into the late-interaction mechanism, further boosting effectiveness. 


\section{Conclusion}

In this work, motivated by insights from interpretability~\cite{zhan2020bertAnalysisReranker,lu-etal-2025-pathway} and preliminary experiments, we proposed MICE, a new architecture that starts bridging the gap between lightweight late-interaction models and heavy cross-encoders. By combining mid-fusion, cross-attention, lexical head, and layer pruning, MICE defines a new effectiveness-efficiency tradeoff.
Our experiments across the BERT and ModernBERT backbones confirm this. MICE reduces computational overhead up to 2.5$\times$ compared to standard cross-encoders, matching late-interaction models like ColBERT while retaining most of cross-encoder ID effectiveness and demonstrating superior generalization abilities.

\section{Limitations}

In this paper, we propose a new architecture for cross-encoders named MICE. We show on different backbones how MICE can significantly improve efficiency while preserving effectiveness. However, our empirical results suggest that there is no single recipe, applicable to any backbone cross-encoder, to turn it into its MICE counterpart: the optimal configuration depends on the differences between backbone encoders such as BERT and ModernBERT. This limits the generalization of our method, and adapting MICE to a new backbone still requires some per-backbone tuning. Relatedly, our evaluation is restricted to English retrieval (MS MARCO and BEIR) and to two encoder families (BERT and ModernBERT) over a limited range of model sizes; we therefore do not provide evidence that our findings transfer to other languages or to substantially different architectures.

Another limitation concerns our evaluation of the OOD abilities of MICE. As we rely on nano-BEIR for validation when training MICE, our results on BEIR may be biased by the small amount of data used to pick the checkpoints across our random seeds (each nano-BEIR dataset is a subset of 5000 documents and 50 queries of the original dataset). This affects our reported OOD numbers, and we cannot rule out that a different validation set would lead to different selected checkpoints. Two factors limit the impact: all of our design choices about MICE, including the masking experiments, were motivated using ID-only evaluations, and the same checkpoint-selection methodology is applied to our baselines and reproductions, so models are compared on equal footing. A more thorough OOD evaluation, e.g.\ adding the LoTTE benchmark \cite{santhanam2022colbertv2}, would have increased the computational cost significantly given the scale of our study (3 random seeds per model $\times$ 26 backbones\footnote{Counting only the ones reported in the \Cref{tab:MICE-evals-full}.} $\times$ 5 more datasets), and we leave it to future work.   

In \Cref{fig:pareto}, we assumed document representations could be pre-computed and stored uncompressed. This is unrealistic at scale, and we do not measure the storage cost or its impact on latency; optimizing multi-vector storage is a separate line of work, out of the scope of this submission, which aims at proposing a new alternative to cross-encoders that allows pre-computing document representations. MICE would likely benefit from such optimization, but our reported efficiency gains should be read under this no-compression assumption.

Finally, MICE remains a re-ranker: our attempts to push it toward first-stage retrieval were unsuccessful. Removing the self-attention over query tokens in the interaction layers collapsed ranking performance, indicating that intra-query contextualization is critical for effective re-ranking. Reducing the dimensionality of the interaction layers proved less effective than layer pruning, as the compressed layers cannot be initialized from the backbone's pre-trained weights. Combining MICE with PLAID \cite{santhanam2022plaid} to compress its pre-computed document embeddings led to a large drop in performance.

\section*{Acknowledgements}
The authors acknowledge the ANR – FRANCE (French National Research Agency) for its financial support of the GUIDANCE project n°ANR-23-IAS1-0003 as well as the Chaire Multi-Modal/LLM ANR Cluster IA ANR-23-IACL-0007. This work was granted access to the HPC resources of IDRIS under the allocations 2025-A0191016944, 2024-AD011015440R1 and 2025-AD011014444R2 made by GENCI. The authors also gratefully acknowledge the support of the Centre National de la Recherche Scientifique (CNRS) through a research delegation awarded to J. Mothe.

\bibliography{biblio}

\appendix

\section{Additional Details}

\subsection{Reproducibility Statement}\label{appendix:Repro_statement}
To facilitate rigorous community evaluation and ensure full reproducibility, we release our complete modular codebase alongside all configuration files required to replicate our training and evaluation pipelines. 

\subsection{Hardware Configuration.} All training experiments are conducted using a single NVIDIA H100 (80GB) GPU. Training is highly efficient with MICE's streamlined architecture: a typical run for a MiniLM-based MICE variant requires approximately three hours, depending on the specific architectural parameters. For inference and large-scale evaluation benchmarks, we utilize NVIDIA V100 (32GB) GPUs. All input tokenization, sequence lengths and evaluation methodology remain constant across these environments to ensure parity.

\subsection{Experimental Setup }\label{appendix:experimental_framework}

\paragraph{Training hyperparameters (Masking experiment, \Cref{sec:masking}).}\label{par:Traindata}

To fine-tune our models and baselines, we rely on distillation with the MarginMSE loss~\cite{hofstatterImprovingEfficientNeural2020}. This loss yields superior retrieval performance compared to the standard Binary Cross Entropy (BCE) \cite{xu2025distillationversuscontrastivelearning}, by preserving the magnitude of the relevance difference (margin) between positive and negative pairs, rather than treating them as binary labels. 
As a teacher, we use the set of re-rankers scores on the MS MARCO passage ranking dataset~\cite{msmarco} from \citeauthor{hofstatterImprovingEfficientNeural2020}, already  used to train efficient re-ranking architectures such as ColBERT \cite{khattab2020colbert}.

We use the same training hyperparameters in all experiments and train all models for 125,000 steps using a batch size of 32, a learning rate of $7\times10^{-6}$, and 5,000 warmup steps. We validate every 10,000 steps based on RR@10 performance on the MS MARCO development set and keep the best checkpoint.

\paragraph{Training hyperparameters (MICE, \Cref{sec:MICE}).}\label{par:TraindataMICE}

The set of hyperparameters used to train our MICE models is the same as the one to learn our masks in \Cref{sec:masking}. The only difference, is in the validation data. Unlike for learning the masks, we follow the \texttt{sentence-transformers} library guidelines \footnote{\href{https://sbert.net/index.html}{https://sbert.net/index.html}} and use nano-BEIR \cite{NanoBEIRMultilingualInformation2026} as a proxy for monitoring convergence. These subsets of the 13 public datasets of BEIR, allow for frequent validation without the overhead of using the full datasets.

\subsection{Models}
Configurations of all the base models used in this work are detailed in \Cref{tab:model-configs}.

\input{new_figures/table_models}

\subsection{Datasets}\label{appendix:datasets}

\input{new_figures/datasets}

\paragraph{Evaluation.}\label{par:evaluation}
We evaluate our models both \emph{in-domain (ID)}, using MS MARCO~\cite{msmarco} dev set (MSM) and the high-quality annotations of the TREC Deep Learning tracks from 2019 (DL19) and 2020 (DL20) \cite{craswell2019overview,craswell2020overview}; and \textit{out-of-domain (OOD)}, where we rely on the publicly available subset of datasets of the BEIR benchmark \cite{thakur_beir_2021} (listed in \Cref{tab:beir_datasets}). This includes the following list of 13 datasets: ArguAna (Ar), Climate-FEVER (CF), DBPedia (DB), FEVER (FE), FiQA (Fi), HotpotQA (HPQ), NFCorpus (NFC), Natural Questions (NQ), Quora Question Pairs (Q), SCIDOCS (SD), SciFact (SF), Touché-2020 (T-v2), and TREC-COVID (T-C).

\section{Masking Experiments}

\subsection{Additional Details on the Masking Steps}\label{appendix:mask_details}

To further motivate the masking steps described in \Cref{sec:masking}, we provide additional details behind their design.

\noindent\textbf{Masking Step 0}

\noindent Across all layers, \SEP{} tokens are prevented from receiving information from any input part ($\SEP \blocked \{\CLS,Q,D\}$), and \CLS{} is prevented from sending information to other parts ($\{Q,\SEP,D\} \blocked \CLS$). We further enforce that \SEP[1] and \SEP[2] act as dedicated attention sinks respectively for $Q$ and $D$ — allowing $Q\allowed$\SEP[1] and $D\allowed$\SEP[2] while blocking $Q\blocked$\SEP[2] and $D\blocked$\SEP[1] — as attention sinks are crucial for absorbing undesirable interactions \cite{zhan2020bertAnalysisReranker}. \\

\noindent\textbf{Masking Step 1}

\noindent The motivation for blocking $\CLS \blocked \{D$,\SEP[2]$\}$ stems from empirical evidence that query tokens act as the primary recipients of query-document interactions, accumulating relevance signals that are subsequently routed to \CLS{} \cite{lu-etal-2025-pathway}. Blocking the direct document-to-\CLS{} path thus removes a redundant and potentially noisy channel. Subsequent ablations (see Appendix~\ref{app:Ablations}), show that blocking this interaction is found to be detrimental to MICE.\\

\noindent\textbf{Masking Step 2}

\noindent Prior work has shown that query-to-document interactions ($D\blocked Q$) contribute less to the overall ranking process than document-to-query ones \cite{zhan2020bertAnalysisReranker}, motivating their removal across all layers. Beyond efficiency, this step paves the way towards architectures where document representations can be computed offline \cite{macavaneyEfficientDocumentReRanking2020}. Note that this step directly contradicts \cite{schlattInvestigatingEffectsSparse2024} decision to mask $Q\blocked D$ in its Sparse cross-encoder design.\\

\noindent\textbf{Masking Step 3}

\noindent The rationale for blocking $Q\blocked D$ in the early layers is that the model primarily contextualizes query and document tokens independently in this stage, with cross-interactions being secondary \cite{zhan2020bertAnalysisReranker,lu-etal-2025-pathway}. We define $\ell^*$ as the highest layer up to which this mask can be applied without degrading the base model's effectiveness, and determine it empirically.

\subsection{Impact of masking superfluous interactions}

To assess how masking superfluous interactions between input parts within the self-attention modules affects cross-encoder effectiveness, we report in \Cref{tab:MskEvals} the results obtained by applying our masks to two off-the-shelf cross-encoder models, based on our two backbones: \texttt{cross-encoder/ms-marco-MiniLM-L12-v2} for MiniLM-v2, and \texttt{tomaarsen/ms-marco-ettin-32m-reranker} for Ettin-32M.

From \Cref{tab:MskEvals}, we observe that the MiniLM-based cross-encoder and the Ettin-based cross-encoder respond very differently to the different masking strategies. 
Although the base effectiveness of the MiniLM model remains stable for masks \ref{mask_step:0} and \ref{mask_step:1} (around 1 nDCG@10 point in average on both ID and OOD), these masks impact severely the cross-encoder based on Ettin (drop of more than 30 nDCG@10 on MS MARCO for \Cref{mask_step:0}). A possible explanation is that our masking steps are derived from \citet{lu-etal-2025-pathway}, who study the MiniLM-v2 cross-encoder model only; they may not fit the internal mechanisms of Ettin-based cross-encoders. For instance on Ettin, while \Cref{mask_step:0} and \Cref{mask_step:1} are intended to target interactions that should have only a marginal effect on model performance (by limiting information flow toward attention sinks, as observed with MiniLM), it is plausible that, because they are based on ModernBERT and pre-trained with mechanisms such as sliding-window attention (unlike BERT-base models), their attention sinks are different from those of BERT-based models. As a result, and given the effect of our masking strategy that redistributes the attention probability that was concentrated on the sink, our masks may be less appropriate for Ettin than for MiniLM. We also observe that \Cref{mask_step:1} increases performance over \Cref{mask_step:0} for Ettin, but it remains much lower than with the unmasked model. 
Finally, while \Cref{mask_step:0} and \Cref{mask_step:1} only slightly impact MiniLM cross-encoder, our results indicate that further masking (\Cref{mask_step:2}) leads to a more substantial decrease of its effectiveness (-10 on nDCG@10 for both ID and OOD). 

\input{new_figures/table_maskedEval}

Together, these results indicate that it is possible to remove some interactions inside the self-attention modules of a cross-encoder, without impacting its effectiveness (see \Cref{mask_step:0} and \Cref{mask_step:1} for MiniLM). However, doing so requires a good understanding of the model internal mechanisms as acknowledged by the results with Ettin. These insights, in addition to the substantial drop induced by \Cref{mask_step:2}, suggest that it is possible to maintain a cross-encoder performance while removing unnecessary interactions only up to a certain point. Removing these interactions on an already fine-tuned model seems to harm its effectiveness, showing that the fine-tuned models still have learned to use some of the information transfers we mask to predict relevance. 
Consequently, empirical evidence from Sparse CE \cite{schlattInvestigatingEffectsSparse2024} shows that fine-tuning can preserve the base model’s effectiveness even when a key information transfer is removed. This naturally motivates fine-tuning the re-ranker with our masks applied, as studied in the main body of the article (see \Cref{sec:masking}).

\input{new_figures/TableMicev2_full}

\section{Ablations on MICE Architectue}\label{app:Ablations}

To validate our architectural choices, we report different ablations of the MICE architecture presented in \Cref{sec:archi}. 
To limit computational overhead, we limit to evaluations on the ID datasets listed in \Cref{appendix:datasets}.

\begin{figure}[t]
    \centering
    \includegraphics[width=\linewidth]{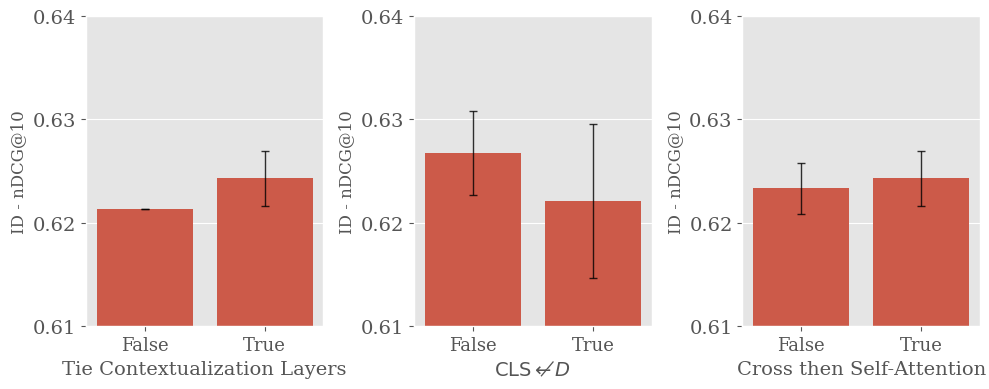}
    \caption{Ablations of three design choices in MICE: 1) whether or not to keep the weights of the contextualization layers tied; whether or not to allow transfer of information from the document to the \CLS{} (see \Cref{mask_step:1}); whether or not to do the cross-attention first or the self-attention first. Impact is measured on the 3 ID datasets (MSM, TREC-DL19 and TREC-DL20).}
    \label{fig:ablation_MICE}
\end{figure}

From \Cref{fig:ablation_MICE} we conclude that:
\begin{itemize}
    \item Unlike MORES \cite{gaoModularizedTransfomerbasedRanking2020mores}, MICE does not draw any benefit from having a separate encoder for the query and the document.
    \item Contrary to our results with \Cref{mask_step:1} in \Cref{sec:masking}, it seems that MICE profits from $\CLS \allowed D$ transfers. While counter-intuitive, we attribute this difference to the fact that in MICE, we also prevent document contextualization in the interaction layers. This change itself might explain that the model needs an additional degree of liberty to predict relevance.
    \item Traditionally, a cross-attention layer in decoders is composed of a self-attention, followed by a cross-attention. Yet, in their design, MORES \cite{gaoModularizedTransfomerbasedRanking2020mores} exchanged these two modules, which prompted us to compare these two variations. Our conclusion confirms their decision. 
\end{itemize}

\section{MICE Full Results}\label{appendix:full_mice_results}

To complement the results from \Cref{tab:MICE-evals}, we detail in \Cref{tab:MICE-evals-full} all the results for all datasets on which MICE models were evaluated. We also include evaluations for MICE models trained from other backbones than MiniLM and Ettin-32M, showing the generalization of our proposed architecture, and further reports the results of ColBERTv2 \cite{santhanam2022colbertv2}. While we do not report any $\text{BERT}_\text{base}$ variant of MICE to directly compare it to ColBERTv2, our results still show that aside from the smaller Ettin-32M, all other variants of MICE match its ID results, while outperforming it with a large margin on BEIR. While we cannot fully discard the role played by the different architectures in this comparison, the fact that MICE variants based on much smaller backbones than ColBERTv2 (MiniLM, or Ettin-68M) manages to exceed its effectiveness underlines that our design brings additional capacity to the model. As a result, MICE can compare favorably with state-of-the-art late-interaction models, both in terms of effectiveness and efficiency. Furthermore, we note that \Cref{tab:MICE-evals-full} results confirm a trend already observed from \Cref{tab:MICE-evals}: when applied to BERT-based cross-encoders, MICE is able to improve the effectiveness in OOD and the efficiency, while the effectiveness tends to drop slightly with Ettin-based models. Results from ELECTRA \cite{clark2020electra} and the other models in the Ettin-suite \cite{ettin}  this observation. We note that, despite contrasted results in terms of effectiveness, MICE variants for Ettin-\{32M-150M\} all profit from massive FLOPs speedups, around twice their respective baseline. This effect can also be observed in the \Cref{fig:pareto}. 

\section{Computing FLOPs in a transformer model}\label{app:flops}

To quantify the computational efficiency of each masking steps and MICE, we estimate the Floating Point Operations (FLOPs) required for a forward pass.
By introducing an interaction mask $M \in \{0, 1\}^{S \times S}$, we \textit{theoretically} reduce this cost proportional to the masking density $\alpha_\ell = \frac{\|M\|_0}{S^2}$. Where $\alpha_\ell=1$ for a standard cross-encoder with full attention.
Let $L$ denote the number of layers, $d$ the model dimension, and $d_{\text{FF}}$ the MLP intermediate dimension. For a sequence of length $S = n + m + 3$ (query length $n$, document length $m$), the FLOPs for a transformer layer with attention density $\alpha_\ell$ is:

\begin{equation}
    C_{\ell} \approx 2 \times S \ ( \underbrace{4 d^2}_{\text{Proj}} + \underbrace{2 d d_{\text{FF}}}_{\text{MLP}} )  + \underbrace{4 \alpha_\ell d_h  \ S^2}_{\text{Attention}}
\end{equation}

where the factor of two comes from the multiply-accumulate operation used in matrix multiplication. 
In \Cref{tab:MICE-evals}, we report a theoretical FLOPs speedup: in practice, we compute it as FLOPs(baseline)  / FLOPs(model) to get the approximate speedup for computing one relevance score. 
We use $S=512$, $n=32$, and $m=S-n-3$ for all models. 
This method yields a 2.5 $\times$ speedup for the MiniLM-based MICE-$\ell$4+3 variant shown in \Cref{tab:MICE-evals}.

\begin{table}[ht]
\centering
\label{tab:layer_flops}
\caption{FLOPs estimates for a single transformer layer. FLOPs are calculated for a full sequence of length $S$. The interaction step incorporates the masking density parameter $\alpha_\ell \in [0, 1]$, where $\alpha_\ell=1$ recovers standard full attention. Sub-leading terms such as nonlinearities, biases, and layer normalization are omitted, following~\citet{kaplan2020scalinglawsneurallanguage}.}
\renewcommand{\arraystretch}{1.4}
\centering
\resizebox{\linewidth}{!}{
\begin{tabular}{lll}
    \hline
    \textbf{Operation} & \textbf{FLOPs ($S$ tokens)} \\
    \hline
    Attention: QKV & $6 S d^2$ \\
    \hline
    Attention: Project & $2 S d^2$ \\
    \hline
    Masked Attention & $4 \alpha_\ell S^2 d$ \\
    \hline
    Feedforward & $4 S d \cdot d_{\text{FF}}$ \\
    \hline
    \textbf{Total} & $ 2 \times S (4d^2 + 2 dd_{\text{FF}}) + 4 S^2 \alpha_\ell  d$ \\
    \hline
\end{tabular}
}
\vspace{0.2cm}

\end{table}

Using this method, our architectural pruning lead to several savings:

\textbf{(1) Block-Diagonal Attention:}
For the first $\ell^*$ layers, query and document are contextualized independently. The attention mask density is reduced to $\alpha_{\text{early}} = \frac{n^2 + m^2}{S^2}$, transforming the quadratic bottleneck from $O(S^2)$ to $O(n^2 + m^2)$ operations.

\textbf{(2) Frozen Document in Interaction Layers:}
In the $L_{\text{int}}$ interaction layers, only query representations are updated. Document tokens are frozen, bypassing their MLP and self-attention computations. The cost per interaction layer is:
\begin{equation}
    C_{\text{int}} \approx \underbrace{8 n d^2 + 4 n d \cdot d_{\text{FF}}}_{\text{Query Projections \& MLP}} + \underbrace{4 n(n+m) d}_{\text{Cross-Attention}}
\end{equation}
This replaces $O(S^2)$ complexity with $O(n \cdot S)$. Since $n \ll m$, this provides a substantial speedup.

\textbf{(3) Layer Pruning:}
MICE discards $L_{\text{drop}}$ layers. The total FLOPs for MICE is:
\begin{equation}
    C_{\text{MICE}} = \ell^* \cdot C_{\ell}(\alpha_{\text{early}}) + L_{\text{int}} \cdot C_{\text{int}}
\end{equation}
where $\ell^* + L_{\text{int}} = L - L_{\text{drop}}$. For the MiniLM MICE-$\ell$4+3 model, this reduces the total active layers from 12 to 7.

\end{document}

%% file: new_figures/Table1.tex
\begin{table}[t]
\centering
\caption{Re-ranking evaluation results over 1k docs/query from BM25 in nDCG@10 (over 5 seeds) for the masking experiment (\Cref{sec:masking}), comparing models fine-tuned with and w/o masking. 
\textbf{Bold} marks the best value per backbone, ``\textunderscore'' second best.}
\vspace{-.3cm}
\resizebox{\linewidth}{!}{
\begin{tabular}{rr|ccc|cc}
\toprule
& &\multicolumn{3}{c}{\textbf{In-domain}} & \multicolumn{2}{c}{\textbf{Average}} \\
\cmidrule(lr){3-5}\cmidrule(lr){6-7}
& \textbf{Re-Ranker} & MSM & DL19 & DL20 & \textbf{ID} & \textbf{BEIR} \\
\midrule
\midrule
\multicolumn{2}{r}{BM25} & 23.0 & 51.2 & 47.7 & 42.5 & 43.0 \\
\midrule
\multirow{7}{*}{\rotatebox{90}{\textbf{MiniLM}}}
& Sparse CE        & 44.1 & \textbf{74.3} & 70.7 & 63.0 & 44.9 \\
& PreTTR ($\ell4$) & 44.5 & \underline{73.8} & 71.9 & 63.4 & \underline{49.2} \\
\cmidrule(lr){2-7}
 & Baseline                   & \underline{44.7} & \underline{73.8} & \underline{72.9} & 63.8 & 44.7 \\
 & \cref{mask_step:0}          & 44.6 & \underline{73.8} & \textbf{73.3} & \underline{63.9} & 46.3 \\
  & \cref{mask_step:1}          & \textbf{44.8} & \underline{73.8} & \textbf{73.3} & \textbf{64.0} & \textbf{50.2} \\
 & \cref{mask_step:2}         & 44.1 & 73.1 & 71.8 & 63.0 & 49.0 \\
 & \cref{mask_step:3}-$\ell$4  & 43.9 & 73.0 & 70.9 & 62.6 & 47.2 \\
\midrule
\midrule
\multirow{5}{*}{\rotatebox{90}{\textbf{Ettin-32M}}} 
 & Baseline                   & \textbf{43.1} & \underline{70.3} & \textbf{69.7} & \textbf{61.0} & \underline{48.3} \\
 & \cref{mask_step:0}         & \underline{42.8} & 70.0 & \textbf{69.7} & 60.8 & 47.8 \\
 & \cref{mask_step:1}         & 42.6 & \textbf{70.8} & \underline{69.2} & \underline{60.9} & \textbf{48.7} \\
 & \cref{mask_step:2}         & 42.3 & 69.6 & 68.5 & 60.1 & 46.7 \\
& \cref{mask_step:3}-$\ell$6  & 41.9 & 69.4 & 68.6 & 59.9 & 42.8 \\
\bottomrule
\end{tabular}
}
\label{tab:MskEvalsFT}
\vspace{-.3cm}
\end{table}

%% file: new_figures/TableMicev2.tex
\begin{table}
\caption{MICE models evaluation of  (re-ranking BM25 top 1K - nDCG@10 over 5 seeds). ``MICE-$\ell$\texttt{X}+[\texttt{Y}/all]'' models indicates using \texttt{Y} interaction layers (or all otherwise) starting from layer \texttt{X}. $\uparrow$ and $\downarrow$ marks a statistically significant difference between models and their cross-encoder baseline. \textbf{Bold} values marks the best averaged value per backbone, ``\textunderscore'' indicates second best. ``$\times$'' column is the FLOPs-based theoretical speedup compared to the baseline (see Appendix~\ref{app:flops} for more details).
}
\label{tab:MICE-evals}
\resizebox{\linewidth}{!}{
\begin{tabular}{lc|ccc|cc}
\toprule
& & \multicolumn{3}{c}{\textbf{In-domain (ID)}} & \multicolumn{2}{c}{\textbf{Average}} \\
\cmidrule(lr){3-5}\cmidrule(lr){6-7}
\quad  \textbf{Re-Ranker} & $\times$ & MSM & DL19 & DL20 & \textbf{ID} & \textbf{BEIR}\\
\toprule
\toprule
\multicolumn{2}{l}{\textbf{MiniLM-L12 - Baseline}} & \textbf{45.0} & \textbf{74.0} & \textbf{73.5} & \textbf{64.1} & 44.0 \\
\quad MORES-$\ell$10-d12+2 & 1.0 & 43.5$\downarrow$ & 72.0$\downarrow$ & 70.1$\downarrow$ & 61.9$\downarrow$ & 48.6$\uparrow$ \\
\quad PreTTR-$\ell$8+4 & 1.0 & \underline{43.8}$\downarrow$ & 71.8$\downarrow$ & 70.9$\downarrow$ & 62.2$\downarrow$ & 46.5$\uparrow$ \\
\quad ColBERT & 1.0 & 38.2$\downarrow$ & 69.1$\downarrow$ & 65.2$\downarrow$ & 57.5$\downarrow$ & 40.6 \\
\cmidrule(lr){1-7}
\quad MICE-$\ell$4+3 & 2.5 & 43.2$\downarrow$ & \underline{73.4} & 70.9$\downarrow$ & \underline{62.5}$\downarrow$ & \underline{49.7}$\uparrow$ \\
\quad MICE-$\ell$4+3 Lex & 2.4 & 43.2$\downarrow$ & 73.1 & 70.5$\downarrow$ & 62.3 $\downarrow$ & \underline{49.7}$\uparrow$ \\
\quad MICE-$\ell$4+all & 1.9 & 43.6$\downarrow$ & 73.3 & 70.8$\downarrow$ & \underline{62.5}$\downarrow$ & \textbf{50.5}$\uparrow$ \\
\quad MICE-$\ell$8+3 & 1.4 & 43.3$\downarrow$ & 70.5$\downarrow$ & 69.7$\downarrow$ & 61.1$\downarrow$ & 49.0$\uparrow$ \\
\quad MICE-$\ell$8+3 Lex & 1.3 & 43.5$\downarrow$ & 72.7$\downarrow$ & \underline{71.2}$\downarrow$ & 62.4$\downarrow$ & \textbf{50.5}$\uparrow$ \\
\midrule
\midrule
\multicolumn{2}{l}{\textbf{Ettin-32M - Baseline}}& \textbf{43.8} & \underline{71.5} & \textbf{71.7} & \textbf{62.4} & \textbf{48.4} \\
\quad MICE-$\ell$3+4 & 2.1 & 42.7$\downarrow$ & 70.5$\downarrow$ & 68.4$\downarrow$ & 60.5$\downarrow$ & \underline{47.9}$\downarrow$ \\
\quad MICE-$\ell$3+4 Lex & 2.0 & 42.4$\downarrow$ & 71.1 & 69.0$\downarrow$ & 60.8$\downarrow$ & 47.6 \\
\quad MICE-$\ell$6+4 & 1.3 & \underline{42.8}$\downarrow$ & \textbf{71.7} & 69.8$\downarrow$ & \underline{61.4}$\downarrow$ & 47.2$\downarrow$ \\
\quad MICE-$\ell$6+4 Lex & 1.2 & 42.6$\downarrow$ & \underline{71.5} & 69.9$\downarrow$ & \underline{61.4}$\downarrow$ & 46.8$\downarrow$ \\
\bottomrule
\end{tabular}
}
\end{table}

%% file: new_figures/table_eff.tex
\begin{table}[!htbp]
\centering
\caption{Efficiency comparison of retrieval models. All based on a \texttt{MiniLM-L12-v2} backbone for comparison.}
\resizebox{\linewidth}{!}{
\begin{tabular}{rrccccc}
\toprule
& \textbf{Model} & \textbf{Precomp.} & \textbf{\#param} & \textbf{Latency (ms)} & \textbf{Docs/s} & \textbf{Peak Mem} \\
\midrule
\multirow{5}{*}{\rotatebox{90}{\textbf{MiniLM}}} 
& MICE $\ell$4+3 & \inlineemoji{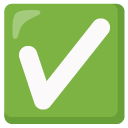} & 26.3M & $113.28 \pm 12.05$ & 1130 & 598.44 MB \\
& ColBERT & \inlineemoji{checkmark} & 33.4M & $130.36 \pm 7.86$ & 982 & 331.77 MB \\
\cmidrule(lr){2-7}
& Cross-Encoder & \inlineemoji{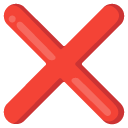} & 33.4M & $470.22 \pm 4.87$ & 267 & 1193.52 MB \\
& MICE $\ell$4+3 & \inlineemoji{cross} & 26.3M & $241.05 \pm 6.25$ & 531 & 1071.61 MB \\
& ColBERT & \inlineemoji{cross} & 33.4M & $498.48 \pm 8.65$ & 257 & 1195.27 MB \\
\bottomrule
\end{tabular}
}
\label{tab:efficiency}
\end{table}

%% file: new_figures/table_models.tex

\begin{table}[ht]
\centering
\small
\caption{Configurations of the models used in this work.}\label{tab:model-configs}
\resizebox{\columnwidth}{!}{
\begin{tabular}{rc}
\toprule
Model & \textbf{MiniLM-L12-v2 (MiniLM)} \\
 Pretrained checkpoint \hf & \texttt{microsoft/MiniLM-L12-H384-uncased}  \\
 Architecture & \textbf{Layers}=12, \textbf{Params}=33M, \textbf{Hidden}=384, \textbf{Heads}=12 \\
\midrule
Model & \textbf{Ettin-32M} \\
 Pretrained checkpoint \hf & \texttt{jhu-clsp/ettin-encoder-32m} \\
 Architecture & \textbf{Layers}=10, \textbf{Params}=32M, \textbf{Hidden}=384, \textbf{Heads}=6 \\
 \midrule
Model  & \textbf{Ettin-68} \\
 Pretrained checkpoint \hf & \texttt{jhu-clsp/ettin-encoder-68m} \\
 Architecture & \textbf{Layers}=19, \textbf{Params}=68M, \textbf{Hidden}=512, \textbf{Heads}=8 \\
\midrule
Model & \textbf{BERT} \\
 Pretrained checkpoint \hf & None (backbone of ColBERTv2) \\
Architecture & \textbf{Layers}=12, \textbf{Params}=110M, \textbf{Hidden}=768, \textbf{Heads}=12 \\
\midrule
Model  & \textbf{ELECTRA} \\
 Pretrained checkpoint \hf & \texttt{google/electra-base-discriminator} \\
Architecture & \textbf{Layers}=12, \textbf{Params}=110M, \textbf{Hidden}=768, \textbf{Heads}=12 \\
\midrule
Model  & \textbf{Ettin-150} \\
 Pretrained checkpoint \hf & \texttt{jhu-clsp/ettin-encoder-150m} \\
Architecture & \textbf{Layers}=22, \textbf{Params}=150M, \textbf{Hidden}=768, \textbf{Heads}=12 \\
\bottomrule
\end{tabular}
}
\end{table}

%% file: new_figures/datasets.tex
\begin{table}[H]
\centering
\caption{List of the 13 publicly available datasets in the BEIR benchmark \cite{thakur_beir_2021}, along with the corresponding abbreviations used in the article and their domain.}
\label{tab:beir_datasets}
\begin{tabular}{llc}
\toprule
\textbf{Dataset} & \textbf{Task} & \textbf{Abb.} \\
\midrule
ArguAna              & Arg. Retrieval   & Ar   \\
Climate-FEVER        & Fact Checking     & CF   \\
DBPedia              & Entity Retrieval  & DB   \\
FEVER                & Fact Checking     & FE   \\
FiQA-2018            & QA  & Fi   \\
HotpotQA             & QA & HPQ  \\
NFCorpus             & Bio-Medical IR              & NFC  \\
Natural Questions    & QA& NQ   \\
Quora & Dup. Questions   & Q    \\
SCIDOCS              & Citation Prediction & SD \\
SciFact              & Fact Checking    & SF   \\
Touché-2020          & Arg. Retrieval   & T-v2 \\
TREC-COVID           & Bio-Medical IR              & T-C  \\
\bottomrule
\end{tabular}
\end{table}

%% file: new_figures/table_maskedEval.tex
\begin{table}[t]
\centering
\vspace{-.2cm}
\caption{Masking off-the-shelf cross-encoders with our approach (\Cref{sec:masking}). Reranking is performed over 1000 documents retrieved by BM25.}
\vspace{-.3cm}
\resizebox{0.90\linewidth}{!}{
\begin{tabular}{rrccc|cc}
\toprule
& &\multicolumn{3}{c}{\textbf{In-domain}} & \multicolumn{2}{c}{\textbf{Average}} \\
\cmidrule(lr){3-5}\cmidrule(lr){6-7}
& \textbf{Re-Ranker} & MSM & DL19 & DL20 & \textbf{ID} & \textbf{BEIR} \\
\midrule
\multirow{4}{*}{\rotatebox{90}{\textbf{MiniLM}}} 
 & Cross-Encoder      & 45.7 & 75.5 & 73.6 & 64.9 & 49.5 \\
 & + \cref{mask_step:0}  & 44.4 & 74.3 & 72.0 & 63.5 & 48.2 \\
 & + \cref{mask_step:1}  & 44.4 & 73.4 & 72.0 & 63.2 & 48.0 \\
 & + \cref{mask_step:2} & 27.8 & 61.5 & 59.8 & 49.7 & 36.6 \\

\midrule
\multirow{4}{*}{\rotatebox{90}{\textbf{Ettin-32M}}} 
 & Cross-Encoder     & 43.7 & 70.8 & 71.3 & 61.9 & 48.1 \\
 & + \cref{mask_step:0} & 12.0 & 29.0 & 29.3 & 23.4 & 7.5 \\
 & + \cref{mask_step:1} & 27.0 & 56.4 & 52.9 & 45.4 & 26.8 \\
 & + \cref{mask_step:2} & 21.3 & 47.8 & 47.8 & 39.0 & 24.0 \\
\bottomrule
\end{tabular}
}

\label{tab:MskEvals}
\end{table}

%% file: new_figures/TableMicev2_full.tex
\begin{table*}[t]
\caption{Full re-ranking evaluation results. All results over 1k docs/query from BM25 in nDCG@10 (over 5 seeds) for the experiments on MICE. ``MICE-$\ell$\texttt{X}+[\texttt{Y}/all]'' models indicates using only \texttt{Y} interaction layers (or all otherwise) starting from layer \texttt{X}. \texttt{X}$^*$ marks a statistically significant difference between models and their cross-encoder baseline. \textbf{Bold} values marks the best averaged value per backbone, ``\textunderscore'' indicates second best. We also report the theoretical speedup ($\times$) comparing the FLOPs used to compute the relevance score for one Query/document pair.
}
\label{tab:MICE-evals-full}
\resizebox{\textwidth}{!}{
\begin{tabular}{lc|ccc|ccccccccccccc|cc}
\toprule
& & \multicolumn{3}{c}{\textbf{In-domain (ID)}} & \multicolumn{13}{c}{\textbf{BEIR 13}} & \multicolumn{2}{c}{\textbf{Average}} \\
\cmidrule(lr){3-5}\cmidrule(lr){6-18}\cmidrule(lr){19-20}
\textbf{BM25 + Re-Ranker} & \textbf{$\times$} & \textbf{MSM} & \textbf{DL19} & \textbf{DL20} & \textbf{Ar} & \textbf{CF} & \textbf{DB} & \textbf{FE} & \textbf{Fi} & \textbf{HPQ} & \textbf{NFC} & \textbf{NQ} & \textbf{Q} & \textbf{SD} & \textbf{SF} & \textbf{T}-v2 & \textbf{T}-C & \textbf{ID} & \textbf{BEIR}\\
\toprule
BM25 only & & 23.0 & 51.2 & 47.7 & 30.0 & 16.5 & 31.8 & 65.1 & 23.6 & 63.3 & 32.2 & 30.6 & 78.9 & 14.0 & 67.9 & 45.4 & 59.5 & 42.5 & 43.0 \\
\midrule
\textbf{BERT - ColBERTv2} &  & 45.0 & 74.6 & 73.4 & 33.8 & 17.0 & 44.3 & 73.0 & 34.2 & 66.2 & 33.4 & 53.9 & 86.3 & 14.1 & 63.8 & 34.2 & 66.7 & 64.3 & 47.8 \\
\midrule
\textbf{MiniLM - Baseline} & 1 & \textbf{45.0} & \textbf{74.0} & \textbf{73.5} & 16.5 & 11.9 & \textbf{46.2} & 74.7 & 32.5 & \textbf{72.9} & 24.8 & \textbf{56.6} & \underline{81.3} & 12.1 & 49.1 & 24.5 & 68.6 & \textbf{64.1} & 44.0 \\
PreTTR-$\ell$8+4 & 1.0 & \underline{43.8}$\downarrow$ & 71.8$\downarrow$ & 70.9$\downarrow$ & 2.6$\downarrow$ & 21.8$\uparrow$ & 43.3$\downarrow$ & 80.8 & 36.0$\uparrow$ & 68.8$\downarrow$ & 33.8$\uparrow$ & \underline{53.9}$\downarrow$ & 78.4 & 16.0$\uparrow$ & 68.8$\uparrow$ & 27.4$\uparrow$ & 72.8$\uparrow$ & 62.2$\downarrow$ & 46.5$\uparrow$ \\
MORES-$\ell$10-d12+2 & 1.0 & 43.5$\downarrow$ & 72.0$\downarrow$ & 70.1$\downarrow$ & 29.1$\uparrow$ & 25.8$\uparrow$ & 41.8$\downarrow$ & 78.8 & 36.3$\uparrow$ & 68.7$\downarrow$ & 34.0$\uparrow$ & 52.6$\downarrow$ & 78.6$\downarrow$ & 15.8$\uparrow$ & 69.7$\uparrow$ & 29.4$\uparrow$ & 71.6$\uparrow$ & 61.9$\downarrow$ & 48.6$\uparrow$ \\
ColBERT & 1.0 & 38.2$\downarrow$ & 69.1$\downarrow$ & 65.2$\downarrow$ & 29.5$\uparrow$ & 15.0 & 35.0$\downarrow$ & 69.0$\downarrow$ & 26.2$\downarrow$ & 52.5$\downarrow$ & 31.0$\uparrow$ & 46.3$\downarrow$ & 59.3$\downarrow$ & 12.3 & 58.6$\uparrow$ & 25.0 & 72.0$\uparrow$ & 57.5$\downarrow$ & 40.6 \\
\cmidrule(lr){2-20}
MICE-$\ell$4+3 & 2.5 & 43.2$\downarrow$ & \underline{73.4} & 70.9$\downarrow$ & \textbf{35.6}$\uparrow$ & 24.6$\uparrow$ & \underline{44.1}$\downarrow$ & 80.0 & 34.5 & 71.0$\downarrow$ & \underline{34.3}$\uparrow$ & 51.9$\downarrow$ & \textbf{81.5} & 16.0$\uparrow$ & 69.2$\uparrow$ & \textbf{31.7}$\uparrow$ & 72.0$\uparrow$ & \underline{62.5}$\downarrow$ & 49.7$\uparrow$ \\
MICE-$\ell$8+3 & 1.4 & 43.3$\downarrow$ & 70.5$\downarrow$ & 69.7$\downarrow$ & 33.4$\uparrow$ & 24.9$\uparrow$ & 41.8$\downarrow$ & 80.1 & 36.3$\uparrow$ & 68.7$\downarrow$ & 33.7$\uparrow$ & 52.5$\downarrow$ & 78.6$\downarrow$ & 15.7$\uparrow$ & 68.9$\uparrow$ & 29.6$\uparrow$ & 72.9$\uparrow$ & 61.1$\downarrow$ & 49.0$\uparrow$ \\
MICE-$\ell$8+3 Lex & 1.3 & 43.5$\downarrow$ & 72.7$\downarrow$ & \underline{71.2}$\downarrow$ & \underline{35.3}$\uparrow$ & \underline{26.8}$\uparrow$ & 42.7$\downarrow$ & \underline{82.0}$\uparrow$ & \textbf{36.8}$\uparrow$ & 71.3$\downarrow$ & \textbf{34.5}$\uparrow$ & 53.4$\downarrow$ & 80.9 & \textbf{16.1}$\uparrow$ & \underline{70.3}$\uparrow$ & 31.4$\uparrow$ & \textbf{75.1}$\uparrow$ & 62.4$\downarrow$ & \textbf{50.5}$\uparrow$ \\
MICE-$\ell$8+all Lex & 1.3 & \underline{43.8}$\downarrow$ & 71.9$\downarrow$ & 70.6$\downarrow$ & 34.9$\uparrow$ & \textbf{27.0}$\uparrow$ & 43.2$\downarrow$ & \textbf{82.3}$\uparrow$ & \underline{36.5}$\uparrow$ & \underline{71.7}$\downarrow$ & \underline{34.3}$\uparrow$ & 53.5$\downarrow$ & 80.4 & \textbf{16.1}$\uparrow$ & \textbf{70.9}$\uparrow$ & \underline{31.5}$\uparrow$ & \underline{74.6}$\uparrow$ & 62.1$\downarrow$ & \textbf{50.5}$\uparrow$ \\
\midrule
\textbf{ELECTRA - Baseline} & 1 & \textbf{46.1} & \textbf{74.7} & \textbf{74.5} & 21.9 & 26.5 & \textbf{47.5} & \textbf{84.1} & \textbf{39.9} & \textbf{74.7} & \textbf{35.6} & \textbf{58.9} & 82.2 & \textbf{17.3} & \underline{71.9} & 27.6 & \textbf{75.1} & \textbf{65.1} & 51.0 \\
MICE-$\ell$8+3 & 1.4 & \underline{44.8}$\downarrow$ & \underline{73.5} & \underline{72.7}$\downarrow$ & \underline{40.2}$\uparrow$ & \textbf{27.9}$\uparrow$ & 45.1$\downarrow$ & 80.5$\downarrow$ & \underline{38.6}$\downarrow$ & 71.6$\downarrow$ & 34.7$\downarrow$ & 55.3$\downarrow$ & \textbf{83.1} & 16.5$\downarrow$ & \textbf{72.0} & \textbf{29.1} & \underline{74.9} & \underline{63.7}$\downarrow$ & \underline{51.5}$\uparrow$ \\
MICE-$\ell$8+3 Lex & 1.3 & \underline{44.8}$\downarrow$ & 73.2$\downarrow$ & \underline{72.7}$\downarrow$ & \textbf{40.8}$\uparrow$ & \underline{27.3}$\uparrow$ & \underline{45.6}$\downarrow$ & \underline{81.2}$\downarrow$ & 38.5 & \underline{73.0}$\downarrow$ & \underline{34.8}$\downarrow$ & \underline{55.7}$\downarrow$ & \textbf{83.1}$\uparrow$ & \underline{16.7}$\downarrow$ & 71.4$\downarrow$ & \underline{29.0} & 73.6$\downarrow$ & 63.6$\downarrow$ & \textbf{51.6}$\uparrow$ \\
\midrule
\textbf{Ettin32 - Baseline} & 1 & \textbf{43.8} & \underline{71.5} & \textbf{71.7} & 10.2 & \textbf{25.0} & \textbf{42.0} & \textbf{83.4} & \textbf{36.6} & \textbf{71.1} & 33.1 & \textbf{53.0} & 81.3 & \textbf{16.0} & 69.9 & \textbf{31.0} & \textbf{76.9} & \textbf{62.4} & \textbf{48.4} \\
MICE-$\ell$3+4 & 2.1 & 42.7$\downarrow$ & 70.5$\downarrow$ & 68.4$\downarrow$ & \underline{31.4}$\uparrow$ & 22.9$\downarrow$ & \underline{40.9} & 73.1$\downarrow$ & \underline{33.9}$\downarrow$ & \underline{69.4}$\downarrow$ & 32.7 & 50.1$\downarrow$ & 81.1 & \underline{15.5} & \underline{70.7} & 25.7 & \underline{74.9}$\downarrow$ & 60.5$\downarrow$ & \underline{47.9}$\downarrow$ \\
MICE-$\ell$3+4 Lex & 2.0 & 42.4$\downarrow$ & 71.1 & 69.0$\downarrow$ & \textbf{31.9}$\uparrow$ & 22.5$\downarrow$ & 40.4$\downarrow$ & 72.1$\downarrow$ & 33.8$\downarrow$ & 69.0$\downarrow$ & 32.5 & 49.8$\downarrow$ & 81.9 & 15.1$\downarrow$ & 70.3 & \underline{26.6} & 73.3$\downarrow$ & 60.8$\downarrow$ & 47.6 \\
MICE-$\ell$6+4 & 1.3 & \underline{42.8}$\downarrow$ & \textbf{71.7} & 69.8$\downarrow$ & 25.0$\uparrow$ & \underline{23.1}$\downarrow$ & 40.2$\downarrow$ & \underline{73.5}$\downarrow$ & 33.4$\downarrow$ & 68.5$\downarrow$ & \textbf{33.2} & \underline{50.9}$\downarrow$ & \underline{82.5} & 15.3$\downarrow$ & \textbf{70.8} & 25.4$\downarrow$ & 72.4$\downarrow$ & \underline{61.4}$\downarrow$ & 47.2$\downarrow$ \\
MICE-$\ell$6+4 Lex & 1.2 & 42.6$\downarrow$ & \underline{71.5} & \underline{69.9}$\downarrow$ & 22.1 & 22.4$\downarrow$ & 39.8$\downarrow$ & 73.2$\downarrow$ & 33.3$\downarrow$ & 69.0$\downarrow$ & \textbf{33.2} & 50.8$\downarrow$ & \textbf{82.9}$\uparrow$ & 15.0$\downarrow$ & 69.4 & 24.2$\downarrow$ & 73.1$\downarrow$ & \underline{61.4}$\downarrow$ & 46.8$\downarrow$ \\
\midrule
\textbf{Ettin68 - Baseline} & 1 & \textbf{45.6} & \textbf{74.4} & \textbf{73.7} & 15.8 & \textbf{27.7} & \textbf{47.0} & \textbf{86.1} & \textbf{40.3} & \textbf{74.0} & \textbf{35.0} & \textbf{57.4} & 81.4 & \textbf{17.2} & \textbf{72.5} & \textbf{31.3} & \textbf{81.8} & \textbf{64.6} & \textbf{51.3} \\
MICE-$\ell$10+4+Lex & 1.6 & 43.8$\downarrow$ & 71.4$\downarrow$ & 72.2$\downarrow$ & \textbf{30.5}$\uparrow$ & 21.6$\downarrow$ & 42.6$\downarrow$ & 75.8$\downarrow$ & 35.8$\downarrow$ & 71.2$\downarrow$ & 33.5$\downarrow$ & 53.5$\downarrow$ & \underline{83.4} & 15.3$\downarrow$ & 70.0 & 27.2$\downarrow$ & 75.6$\downarrow$ & 62.5$\downarrow$ & 48.9$\downarrow$ \\
MICE-$\ell$8+3 & 2.1 & \underline{44.7}$\downarrow$ & 72.0$\downarrow$ & \underline{72.5}$\downarrow$ & 28.0$\uparrow$ & \underline{24.3}$\downarrow$ & \underline{44.5}$\downarrow$ & 76.9$\downarrow$ & \underline{36.2}$\downarrow$ & 71.0$\downarrow$ & \underline{34.6}$\downarrow$ & \underline{54.0}$\downarrow$ & \textbf{83.7} & \underline{16.2}$\downarrow$ & 70.8 & 26.7$\downarrow$ & \underline{77.3}$\downarrow$ & \underline{63.1}$\downarrow$ & 49.6$\downarrow$ \\
MICE-$\ell$8+3+Lex & 2.0 & 44.5$\downarrow$ & \underline{72.3}$\downarrow$ & 72.2$\downarrow$ & \underline{29.9}$\uparrow$ & 23.8$\downarrow$ & 44.1$\downarrow$ & \underline{77.9}$\downarrow$ & 36.0$\downarrow$ & \underline{71.3}$\downarrow$ & 34.4$\downarrow$ & 53.8$\downarrow$ & \underline{83.4} & \underline{16.2}$\downarrow$ & \underline{71.3} & \underline{28.0} & \underline{77.3}$\downarrow$ & 63.0$\downarrow$ & \underline{49.8}$\downarrow$ \\
\midrule
\textbf{Ettin150 - Baseline} & 1 & \textbf{46.2} & \textbf{74.8} & \textbf{74.0} & 21.9 & \textbf{28.3} & \textbf{47.7} & \textbf{85.0} & \textbf{40.9} & \textbf{75.4} & \textbf{35.9} & \textbf{59.0} & 77.4 & \textbf{18.3} & \textbf{74.9} & \textbf{29.2} & \textbf{82.2} & \textbf{65.0} & \textbf{52.0} \\
MICE-$\ell$12+3 & 1.7 & 45.5$\downarrow$ & \underline{74.6} & \underline{73.7} & \textbf{39.8}$\uparrow$ & \underline{23.9}$\downarrow$ & \underline{46.4}$\downarrow$ & 78.7$\downarrow$ & \underline{37.7}$\downarrow$ & 73.4$\downarrow$ & \underline{35.2}$\downarrow$ & 56.6$\downarrow$ & \underline{83.9} & 16.8$\downarrow$ & \underline{74.2} & 27.1$\downarrow$ & 79.9$\downarrow$ & \underline{64.6} & \underline{51.8} \\
MICE-$\ell$12+4 & 1.5 & \underline{45.6}$\downarrow$ & 74.3 & \underline{73.7} & \underline{35.3}$\uparrow$ & 23.6$\downarrow$ & \underline{46.4}$\downarrow$ & \underline{78.8}$\downarrow$ & 37.5$\downarrow$ & \underline{74.1}$\downarrow$ & \underline{35.2}$\downarrow$ & \underline{56.8}$\downarrow$ & \textbf{84.7} & \underline{17.0}$\downarrow$ & 73.6 & \underline{27.4} & \underline{80.5} & 64.5$\downarrow$ & 51.6 \\
\bottomrule
\end{tabular}
}
\end{table*}